\documentclass[a4paper,11pt]{article}
\pdfoutput=1 

\usepackage{jheppub} 

\usepackage{color,amsthm,amsmath,amsfonts,graphicx,bm,amssymb,hyperref, tikz, pgfplots, braket}
\usepackage{epstopdf}
\usepackage{comment}
\usepackage{etoolbox}
\usepackage{marginnote}
\usepackage{subfigure}
\usepackage{geometry}
\usepackage{booktabs}
\usepackage{array}

\title{Chiral entanglement in massive quantum field theories in 1+1 dimensions}

\author[a]{M. Lencs\'es,} \author[a,b]{J. Viti} \author[c]{ and G. Tak\'acs}

\affiliation[a]{International Institute of Physics, UFRN\\
Campos Universit\'ario, Lagoa Nova 59078-970 Natal, Brazil}
\affiliation[b]{Escola de Ci\^encia e Tecnologia, UFRN\\ 
Campos Universit\'ario, Lagoa Nova 59078-970 Natal, Brazil}
\affiliation[c]{BME \lq\lq Momentum\rq\rq Statistical Physics Research Group\\
Department of Theoretical Physics, Budapest University of Technology and Economics\\ 1111 Budapest, Budafoki \'ut 8, Hungary}

\emailAdd{mate.lencses@gmail.com}
\emailAdd{viti.jacopo@gmail.com}
\emailAdd{takacsg@eik.bme.hu}

\date{11th November 2018}

\abstract{We determine both analytically and numerically the entanglement between chiral degrees of freedom in the ground state of massive perturbations of 1+1 dimensional conformal field theories quantised on a cylinder. Analytic predictions are obtained from a variational Ansatz for the ground state in terms of smeared conformal boundary states recently proposed by J. Cardy, which is validated by numerical results from the Truncated Conformal Space Approach. We also extend the scope of the Ansatz by resolving ground state degeneracies exploiting the operator product expansion. The chiral entanglement entropy is computed both analytically and numerically as a function of the volume. The excellent agreement between the analytic and numerical results provides further validation for Cardy's Ansatz. The chiral entanglement entropy contains a universal $O(1)$ term $\gamma$ for which an exact analytic result is obtained, and which can distinguish energetically degenerate ground states of  gapped systems  in 1+1 dimensions. }

\begin{document}
\maketitle
\flushbottom

\section{Introduction}
\label{sec:intro}

Entanglement entropy plays a central role both in Quantum Field Theory (QFT) and condensed matter theory~\cite{AFO, Eis_rev, Nis_rev, Laflor_rev}. Universal behaviour of entanglement can be used to deduce information about the conformal field theory underlying quantum phase transitions~\cite{HLW, VLRK, CC, CCcft}, while studying the entanglement spectrum leads to deeper insights of topological properties~\cite{KP,LevinWen} of quantum Hall states~\cite{LH}. In non-equilibrium situations entanglement and entropy production are deeply connected \cite{CC2, KTLRSPG, AC}, and are also diagnostic of non-perturbative effects such as confinement \cite{KCTP} and thresholds in the quasi-particle spectrum~\cite{DynGibbs1, DynGibbs2}. Black hole entropy was identified with the entanglement of modes across the horizon~\cite{BKLS, Srednicki, DasBH}, and entanglement entropy was also connected to gravitational space-time geometry in a holographic context~\cite{RT, NRT}. 

Most of the above developments concern entanglement between spatially separated subsystems. In contrast, the present work proposes a characterisation of the Renormalisation Group (RG) flow in 1+1 dimensional quantum field theories through a quantification of the entanglement between left and right moving excitations, which are decoupled in the ultraviolet  limit and become entangled during the flow to the infrared. 

Non-trivial fixed points of RG~\cite{RG}  trajectories in QFT  are characterised by scale invariance and consequently a gapless spectrum in infinite volume.  In relativistic quantum field theories, scale invariance is generally promoted to conformal symmetry \cite{Polyakov, Polchinski}, which is extended to an infinite dimensional symmetry in the 1+1 dimensional case~\cite{BPZ}. 1+1 dimensional Conformal Field Theories (CFTs) obey holomorphic (chiral) factorisation: the excitation spectrum consists of left  $(l)$ and right $(r)$ movers which transform under a separate chiral symmetry algebra and do not interact. When quantised on a cylinder of circumference $L$, the eigenstates of the CFT Hamiltonian span irreducible representations of the tensor product of two identical Virasoro algebras, one for each chirality~\cite{BPZ}. The Hilbert space of the CFT is further restricted by physical requirements such as modular invariance on the torus~\cite{Cardy_mod}. In this paper we only consider the simplest case when only left and right movers coming from the same representation are paired in the tensor product (the so-called diagonal or more precisely $A$-type modular invariant theories~\cite{CIZ}).

Coupling the CFT  to a relevant scalar field $\phi(x,y)$ with coupling constant $\lambda$ according to the formal Euclidean action
\begin{equation}
\label{act}
\mathcal{A}=
\mathcal{A}_{CFT}+\lambda\int_{\mathbb R} dx\int_{0}^{L} dy~\phi(x,y),
\end{equation}
results in breaking conformal invariance. Here we consider the case when this leads to a finite mass gap $m$, related to the coupling $\lambda$ as 
\begin{equation}
m\propto\lambda^{\frac{1}{2-\Delta_\phi}}
\end{equation}
where $\Delta_{\phi}<2$ is the scaling dimension of the field $\phi$ at the RG fixed point associated to the CFT. The action \eqref{act} describes an RG flow from an ultraviolet (UV) massless fixed point for $mL\ll 1$  to a (trivial) infrared (IR) fixed point  when $mL\gg 1$. In general, the theory is regularised using a high energy (equivalently short distance) cut-off $\Lambda$ which is sent to infinity to 
obtain physical predictions.

Properties of RG flows in 1+1 dimensions have been extensively studied with sophisticated analytical and numerical techniques, inspired by the seminal contributions~\cite{Zam_RG, ZamPotts, YurZam}, with most of the analysis focused on the evolution of the QFT  spectrum.  Real space entanglement was also computed for integrable flows using exact form factors of branch point twist fields \cite{FFentanglement,FFentanglement_review, CastroA_E8, FFentanglement_new} and was studied numerically in~\cite{Palmai}.

However, universal information in the IR limit can also be obtained  determining overlaps between eigenstates of the  QFT Hamiltonian associated to \eqref{act}  and states of the UV conformal basis \cite{GaiottoDW, Kon}. More explicitly, assume that $|\Psi_{\Lambda}(L)\rangle$ is the  ground state of the theory \eqref{act} with a finite cut-off, and define the left/right chiral Hilbert spaces 
\begin{equation}
\mathcal{H}_{l}=\mathcal{H}_{r}=\bigoplus\limits_a \mathcal{V}_a
\end{equation}
as a direct sum  of inequivalent irreducible representations $\mathcal{V}_a$ of the symmetry chiral algebra. Then the Hilbert space of the CFT corresponding to a diagonal modular invariant is the diagonal subspace 
\begin{equation}
\label{conf_Hilbert}
\mathcal{H}_{CFT}=\bigoplus\limits_a \mathcal{V}_a\otimes \mathcal{V}_a
\end{equation}
of $\mathcal{H}_{l}\otimes\mathcal{H}_{r} $. The finite volume ground state of the perturbed conformal field theory can be fully characterised by the overlaps
\begin{equation}
\label{over}
v_{ab}(L)\equiv\lim_{\Lambda\rightarrow\infty}\langle\Psi_{\Lambda}(L)|\cdot\left(| a\rangle_l\otimes |b\rangle_r\right)
\end{equation}
where  $|a\rangle_l$ and $|b\rangle_r$ enumerate two orthonormal basis of $\mathcal{H}_{l}$ and $\mathcal{H}_{r}$. Note that the overlaps \eqref{over} vanish when the left and right chiral vectors transform under different irreducible representations of the chiral algebra. 

Based on the observation that $v_{ab}(L)$ can be considered as a density matrix on the chiral space of states, the \emph{right} chiral reduced density matrix $\rho_r$ can be constructed tracing out the \emph{left} chiral sector with matrix elements 
\begin{equation}
\label{red}
[\rho_r(L)]_{bb'}=\sum_{a} v^{*}_{ab}(L)v_{ab'}(L)
\end{equation}
The spectrum of the right reduced density matrix $\rho_r$ (or equivalently the left $\rho_l$) quantifies the entanglement of the UV chiral degrees of freedom  along the RG flow. We refer to the corresponding von Neumann entropy
\begin{equation}
  \mathcal{S}_{\chiß}\equiv-\text{Tr}[\rho_r\log(\rho_r)]=-\text{Tr}[\rho_l\log(\rho_l)]
\end{equation}
as the \emph{chiral entanglement entropy}.

The chiral entanglement entropy vanishes at the UV fixed point and remains finite along the RG flow; i.e. the limit  $\Lambda\rightarrow\infty$ in Eq.~\eqref{over} always exists. We will demonstrate that for $mL\gg 1$ the chiral entanglement entropy grows linearly with $L$ with a non-universal  (i.e. mass dependent)  slope $\mathcal{B}$ 
\begin{equation}
\label{ent_sum}
\mathcal{S}_{\chi}(L)\sim \mathcal{B}L-\gamma+O\left(e^{-L}\right)\,.
\end{equation}
Moreover, we conjecture that  the  sub-leading term $\gamma$, which is $O(1)$ in the system size, is  universal and provides an independent characterisation of the ground state of the (infinite volume) massive theory, which is especially useful when it is degenerate. Such a term first appeared in the analysis of chiral entanglement of regularised conformal boundary states in~\cite{KL,DubailReadRezayi} (see also~\cite{LREEfreebos,DD,LREEDp}), and was suggested as a possible benchmark for topological states in two spatial dimensions following~\cite{KP,LevinWen}. These claims were tested numerically using matrix product states in~\cite{MPS1,MPS2}.

The idea of ~\cite{KP,LevinWen} regarding the universality of  the $O(1)$ term $\gamma$ can be extended to the ground state of massive perturbations of conformal field theories based on a variational Ansatz in finite volume proposed  recently by J. Cardy~\cite{Cardy17}, which describes the exact ground state  
\begin{equation}
|\Psi(L)\rangle\equiv\lim_{\Lambda\rightarrow\infty}|\Psi_{\Lambda}(L)\rangle
\end{equation}
in terms of smeared conformal boundary states. 

Here we use this Ansatz to determine analytically both the reduced chiral density matrix and its chiral entanglement entropy. In cases where the variational Ansatz predicts degenerate ground states, we resolved the degeneracies utilising the conformal OPEs. As argued in~\cite{Kon} and proved there for free Ising Field Theory, a representation of the ground state of the massive QFT  as a conformal boundary state should be exact in the IR limit $mL\rightarrow\infty$. Based on this assumption, of which we  provide numerical tests in the rest of the paper, Cardy's variational Ansatz should be exact (apart from a divergent normalisation) in infinite volume. This implies that the $O(1)$ term of the chiral entanglement entropy obtained from such an Ansatz in the limit $mL\rightarrow\infty$ is exact and indeed a universal feature of the RG flow. In addition, the Ansatz and the results derived for the chiral entanglement entropy do not depend on any special properties of the flow such as integrability at all. To provide numerical support and verification of our claims, we use the Truncated Conformal Space Approach (TCSA) introduced in \cite{YurZam} (cf.~\cite{JKLRT} for a recent review). First we validate the variational Ansatz, using the scaling Ising Field Theory (IFT) and Tricritical Ising Field Theory (TIFT) as examples, and then analyse the overlaps with the conformal states and the large volume behaviour of the chiral entanglement entropy. 

There are some important differences between the properties of the chiral and real space entanglements. Strictly speaking, the real space ground state entanglement entropy of 1+1 dimensional QFTs does not  contain any universal term. Indeed, the leading term is logarithmically divergent~\cite{HLW} in the limit $\Lambda\rightarrow \infty$ and requires a regularisation, e.g. through introduction of a relative entropy~\cite{CH04, Cas08, L14,Wrev}. For a spatial interval of length $R$ embedded into an infinite system, the coefficient of the leading cut-off dependence of real space entanglement entropy is universal for both $mR\rightarrow 0$ and $mR\rightarrow \infty$, and related to the central charge of the UV theory~\cite{CC, FFentanglement}. This feature has found important applications in statistical mechanics models such as spin systems~\cite{VLRK, JK} where the lattice spacing provides a natural regularisation. However, since the leading cut-off dependence is logarithmic, there cannot be  any universal $O(1)$ term in contrast to $\gamma$ defined from the chiral entanglement entropy in Eq.~\eqref{ent_sum}.

The real space entanglement spectrum was also analysed recently in~\cite{lr}. For a bipartition in two semi-infinite halves and in the limit of a large mass gap, the entanglement Hamiltonian~\cite{BW} is the generator of translation around an annulus with certain conformal boundary conditions~\cite{Cardy89}. Similarly to the entanglement entropy, the entanglement spectrum also exhibits universal logarithmic short-distance cut-off dependence, but it is not finite in the limit $\Lambda\rightarrow\infty$. In contrast, the chiral entanglement spectrum is finite for $\Lambda\rightarrow\infty$; we return to comment on this issue in Section~\ref{sec:ent}.

We finally note that the chiral entanglement entropy can be calculated from TCSA in a rather simple and straightforward way, in marked contrast to the real space entanglement entropy~\cite{Palmai}.  

The outline of the paper is as follows. In Section~\ref{sec:cardy} we review and slightly extend Cardy's variational Ansatz for the QFT ground state in finite volume. In Section~\ref{sec:ent} the variational ground state is used to compute the chiral entanglement entropy and in particular the universal $O(1)$ term characterising the RG flow. In Section~\ref{sec:exmp} we discuss applications to Ising field theory and the Tricritical Ising field theory.  In Section~\ref{sec:TCSA} all our claims contained in the previous sections are subjected to numerical tests based on TCSA. Our conclusions are presented in Section~\ref{sec:conc}. In order to keep the main line of the argument focused, technical details regarding the implementation of numerical computations were relegated to two Appendices.

\section{Variational Ansatz for the ground state of the perturbed CFT}\label{sec:cardy}
The QFT Hamiltonian on the cylinder associated to the Euclidean action in Eq.~\eqref{act} is
\begin{equation}
\label{gs_pert}
H=H_{\text{CFT}}+\lambda\int_{0}^{L} dy~\phi(0,y)\,,
\end{equation}
where the relevant scalar field $\phi$ has conformal  dimension $\Delta_{\phi}\equiv 2h_{\phi}$ at the UV fixed point (see also Eq.~\eqref{hamCFT}).
 The IR  regime is reached for $mL\gg 1$,  which is equivalent to either increasing the volume $L$ or the coupling constant $\lambda$. The UV fixed point is a CFT  with central charge $c$  and a torus partition function which is assumed to correspond to a diagonal modular invariant.  In finite volume any eigenstate of the QFT Hamiltonian can be expanded in the eigenstates of $H_{\text{CFT}}$ which form the so-called conformal basis (cf.~Appendix~\ref{appA}).
\subsection{The Ansatz for the non-degenerate case}
\begin{figure}[t]
\centering
\includegraphics{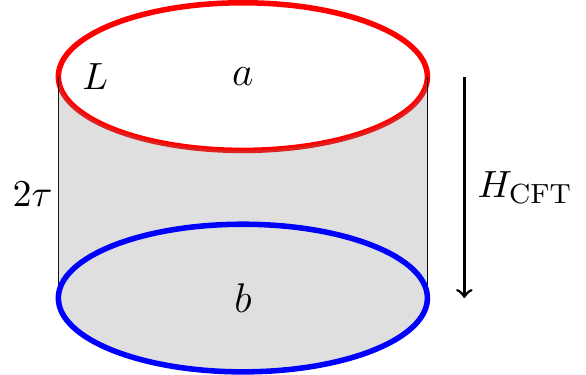}
\caption{Conformal partition function $\mathcal{Z}_{ab}$ on an annulus. The system is quantised on a cylinder of length $L$, the smearing parameter $\tau$ has the dimension of the inverse of the mass gap (correlation length). The conformal Hamiltonian $H_{\text{CFT}}$ generates translation along the annulus, between two boundary states $a$ and $b$.}
\label{fig:ent}
\end{figure}

If we now imagine to turn on the relevant perturbation only for negative imaginary times $x<0$, in the limit $mL\rightarrow\infty$, the perturbation flows in the far infrared to a conformal invariant boundary condition for the theory in the upper half~\cite{CCQuench, Kon, DubailReadRezayi}. In the crossed channel, the conformal boundary condition is represented by a boundary state, while evolution in imaginary time asymptotically projects any state of the QFT onto the ground state of the Hamiltonian. Therefore for a large but finite $mL$ the ground state of the massive QFT is expected to correspond to a deformation of the asymptotic conformal boundary state by boundary irrelevant operators~\cite{DubailReadRezayi}. The simplest such operator is given by the stress tensor $T_{00}$, and Cardy suggested in~\cite{Cardy17}  that it should be possible to approximate the exact ground state of \eqref{gs_pert}  by a suitable linear combination of smeared conformal boundary states 
\begin{equation}
\label{Ansatz}
|\{\alpha_a\},\tau\rangle=\sum_{a}\alpha_{a}e^{-\tau H_{\text{CFT}}}|a\rangle.
\end{equation}
where the $|a\rangle$ are conformal boundary states (defined in detail later) and $\tau$ is a smearing parameter proportional to the inverse of the mass gap. The normalisation of the smeared boundary states is given by
\begin{equation}
\label{part}
\mathcal{Z}_{aa}=\langle a|e^{-2\tau H_{CFT}}|a\rangle,
\end{equation}
which is the conformal partition function on the annulus of Figure~\ref{fig:ent} with boundary conditions $a$ on both sides. In general we will denote by $\mathcal{Z}_{ab}$ the same conformal partition function with boundary condition $a$ and $b$.

The coefficients of the linear combination in Eq.~\eqref{Ansatz} and the smearing parameter are determined by minimising the energy density in infinite volume
\begin{equation}
\label{Cardy_var}
 \mathcal{E}[\{\alpha_a\},\tau]=\lim_{L\rightarrow\infty}\frac{1}{L}\frac{\langle\{\alpha_a\},\tau|H|\{\alpha_a\},\tau\rangle}{\langle\{\alpha_a\},\tau|\{\alpha_a\},\tau\rangle}.
\end{equation}
For $L\gg\tau$, the matrix elements of $H$ are diagonal on the smeared boundary states up to exponentially small corrections~\cite{Cardy17}. In particular it turns out that
\begin{equation}
\label{varE}
\mathcal{E}[\{\alpha_a\},~\tau]=\frac{\sum_{a}\alpha_a^2 e_a(\tau)}{\sum_{a}\alpha_a^2}\,,
\end{equation}
where the functions $e_a$ are given by
\begin{equation}
\label{vare}
e_a(\tau)=\frac{\pi c}{24 (2\tau)^2}+\lambda\tilde{A}^a_{\phi}\left(\frac{\pi}{4\tau}\right)^{\Delta_{\phi}}.
\end{equation}
The coefficients $\tilde{A}^a_{\phi}$ are fixed by the normalisation of the one-point function of the field $\phi$ on the conformal upper half plane~\cite{CL} with conformal boundary condition $a$ on the real axis (see also Eq.~\eqref{ecoff})
\begin{equation}
\label{uhp}
\langle\phi(z)\rangle_{\text{UHP}}=\frac{\tilde{A}_{\phi}^a}{[2\Im m(z)]^{\Delta_{\phi}}}.
\end{equation}
Eq.~\eqref{varE} is minimized keeping in Eq.~\eqref{Ansatz}  only the boundary state $|a^*\rangle$ such that  $e_{a^{*}}(\tau)$ has a global minimum at $\tau=\tau^*$ which is smaller than all the other minima (if any) of the functions $e_a(\tau)$ in Eq.~\eqref{vare}.
As discussed in~\cite{Cardy17}, in the case of a single relevant perturbation with the sign of $\lambda$ positive (negative), the variational Ansatz  selects  $a^*$ such that  $\tilde{A}^{a^*}_{\phi}$ is minimum (maximum). If there is a unique solution of the variational equations, the best approximation of the QFT ground state $|\Psi_\text{var}(L)\rangle$ is then obtained normalising the unique smeared boundary state selected by the variational principle, i.e.
\begin{equation}
\label{Ansatz_f}
|\Psi_\text{var}(L)\rangle=\frac{e^{-\tau^* H_{\text{CFT}}}}{\sqrt{\mathcal{Z}_{a^*a^*}}}|a^*\rangle\,.
\end{equation}
We return to the case of degenerate solutions at the end of the this section.

For perturbations with several relevant fields $\phi_j$  with couplings $\lambda_j$ and $0<\Delta_{\phi_j}<2$ one must find the minimum of 
\begin{equation}
\label{Cardy_bulk}
e_a(\tau)=\frac{\pi c}{24 (2\tau)^2}+\sum_{j}\lambda_j\tilde{A}^a_{\phi_j}\left(\frac{\pi}{4\tau}\right)^{\Delta_{\phi_j}}\,;
\end{equation}
however we will not consider this case here. 

For an explicit evaluation of the variational functions \eqref{vare}, we recall some basic properties of the conformal boundary states in CFTs with diagonal modular invariant partition functions. They can be expressed as a linear combination of Ishibashi states
\begin{equation}
\label{CS}
|a\rangle=\sum_{h}c_{ah}|\Phi_h\rangle\rangle\,,
\end{equation}
where the sum runs  over all allowed irreducible representations of the Virasoro algebra.  The Ishibashi states $|\Phi_{h}\rangle\rangle$ themselves are in one-to-one correspondence with irreducible representations and have the form
\begin{equation}
\label{Ishi}
|\Phi_{h}\rangle\rangle=\sum_{N}\sum_{k}|h,N,k\rangle_{l}\otimes|h,N,k\rangle_{r}\,,
\end{equation}
where the vectors $|h,N,k\rangle$ form an orthonormal basis of the irreducible Virasoro representation of highest weight $h$, with $N$ denoting their descendant level and $k$ a further index enumerating linearly independent vectors at a fixed level. The character of such a representation is given by 
\begin{equation}
\chi_h(q)=\sum_{N,k}q^{h+N-c/24}\quad\textrm{with}\quad q=e^{-\frac{4\pi\tau}{L}}\,.
\label{Virchar}\end{equation}
For a diagonal CFT, the physical conformal boundary states (also called Cardy states) satisfy further constraints~\cite{Cardy89} and are in one-to-one correspondence with the scalar primary fields $\Phi_{h}$  generating the operator content of the theory. Introducing the chiral weight of the primary field  $\Phi_{h_a}$ as $h_a$, the corresponding conformal boundary state is denoted as $|\tilde{h}_a\rangle$. For a Cardy state $|\tilde{h}_a\rangle$ the coefficients of the linear combination in Eq.~\eqref{CS} are 
 \begin{equation}
 c_{ah}=\frac{S^{h}_{h_a}}{[S^{h}_0]^{1/2}}
\end{equation}
where $S$  is the orthogonal symmetric modular $S$ matrix~\cite{Verlinde}, which describes the linear transformation of Virasoro characters
\begin{equation}
\chi_{h}(q)=\sum_{h'}S^{h'}_{h} \chi_{h'}(\tilde{q})
\label{modular_transf}\end{equation}
 under the modular transformation
\begin{equation}
q\rightarrow\tilde{q}=e^{-\frac{L\pi}{2\tau}}\,.
\end{equation}
Then the normalisation coefficient in Eq.~\eqref{uhp} can also be expressed in terms of the modular $S$ matrix as~\cite{CL}
\begin{equation}
\label{ecoff}
\tilde{A}^{a}_{\phi}=\frac{S_{h_a}^{h_{\phi}}}{S_{h_a}^0}\left(\frac{S_0^0}{S_0^{h_{\phi}}}\right)^{1/2}\,,
\end{equation}
which can be inserted into the variational functions \eqref{vare} to obtain them explicitly in terms of  the modular $S$ matrix.

\subsection{The degenerate case}\label{subsec:degenerate}
So far it was assumed that one of the  variational functions \eqref{vare} had a minimum lower than all the others. Suppose that, on the contrary, the minimisation of Eq.~\eqref{vare} leads to two different Cardy states $|\tilde{h}_{a^*}\rangle$ and $|\tilde{h}_{b^*}\rangle$ with $a^*\not=b^*$ such that $e_{a^*}(\tau)$ and $e_{b^*}(\tau)$ have the the same minimum at $\tau=\tau^*$, which can happen e.g. when the two functions $e_{a^*}(\tau)$ and $e_{b^*}(\tau)$ are identical. 

To resolve this problem, note that while the functional \eqref{varE} is obtained for $L\rightarrow\infty$, the degeneracy between candidate smeared conformal boundary states is generally lifted in finite volume. The way the degeneracy is lifted is constrained by the block diagonal structure of the QFT Hamiltonian dictated by the OPE of the perturbing field with the primaries belonging to the CFT. The Hilbert space $\mathcal{H}_{\text{QFT}}$ spanned by the eigenstates of $H$ in Eq.~\eqref{gs_pert} (see also Eq.~\eqref{matH}) then splits into a direct sum
\begin{equation}
\label{QFT_dec}
 \mathcal{H}_{\text{QFT}}=\bigoplus_{\nu} \mathcal{H}_{\text{QFT}}^{\nu}.
\end{equation}
The vector spaces $\mathcal{H}_{\text{QFT}}^{\nu}$ are direct sums of tensor products of left/right  irreducible Virasoro representations as in Eq.~\eqref{conf_Hilbert}. They are composed of modules corresponding to sets of scalar primaries which form a closed set under operator product with the the perturbing field $\phi$ in Eq.~\eqref{act}. The decomposition \eqref{QFT_dec} follows from the explicit expression of the interaction matrix $\text{B}_{ij}$ in Eq.~\eqref{matB} and reflects residual discrete symmetries of the perturbed CFT.

In all the cases considered in this paper the index $\nu$ can have at most two values. For the degenerate case we denote them as $\pm$ and the Hilbert space $\mathcal{H}_{\text{QFT}}^{+}$ (resp. $\mathcal{H}_{\text{QFT}}^{-}$) is identified to be the sector which contains (resp. does not contain) the highest weight representation associated to the identity field. Provided there is no level crossing involving the ground state in finite volume (which is generally the case) the exact QFT ground state belongs to the subspace $\mathcal{H}_{\text{QFT}}^{+}$, whereas in general the degenerate variational states selected by minimizing Eq.~\eqref{vare} do not. 
Consistency requires that the best approximation of the ground state must be the smeared boundary state constructed from a linear combination of the degenerate Cardy states belonging to $\mathcal{H}_{\text{QFT}}^{+}$. It is also possible to construct a linear combination of the degenerate Cardy states  belonging to $\mathcal{H}_{\text{QFT}}^{-}$, which gives the best approximation of a state with an energy gap that decays exponentially with the volume of the system.

Finally we note that the linear combinations of smeared Cardy states obtained from the argument above should be reproduced directly analysing finite size effects in the QFT Hamiltonian restricted to the degenerate subspace spanned by the variational states. If $|\tilde{h}_{a^*}\rangle$ and $|\tilde{h}_{b^*}\rangle$ are the degenerate Cardy states, one must examine the $2\times 2$ matrix
\begin{equation}
\label{matabfull}
\mathcal{M}_{ab}=\frac{1}{\sqrt{\mathcal{Z}_{aa}\mathcal{Z}_{bb}}}\langle \tilde{h}_a | e^{-\tau^* H_{\text{CFT}}}He^{-\tau^* H_{\text{CFT}}}|\tilde{h}_{b}\rangle\,,
\end{equation}
with $H$ given in \eqref{gs_pert} and $a,b=\{a^*, b^*\}$. The true ground state corresponds to the eigenvector of \eqref{matabfull} with the lower eigenvalue in the limit $L\gg\tau^*$ and  belongs to $\mathcal{H}^{+}_{\text{QFT}}$, while the other eigenstate of \eqref{matabfull} is in $\mathcal{H}^{-}_{\text{QFT}}$. Constructing the above matrix is quite involved due to the  matrix elements of the perturbing field $\phi$ which also require the bulk-boundary and boundary OPE structure constants calculated in~\cite{Lew, Run}. However, one can consider a simplified version of this computation by neglecting the perturbing field and replacing $H$ by $H_{\text{CFT}}$, resulting in the matrix
\begin{equation}
\label{matab}
\mathcal{M}_{ab}^0=\frac{1}{\sqrt{\mathcal{Z}_{aa}\mathcal{Z}_{bb}}}\langle \tilde{h}_a | e^{-\tau^* H_{\text{CFT}}}H_{\text{CFT}}e^{-\tau^* H_{\text{CFT}}}|\tilde{h}_{b}\rangle\,,
\end{equation}
which has the leading large volume behaviour
\begin{equation}
\label{matableading}
\mathcal{M}_{ab}^0=L\frac{\pi (c-24h_{ab})}{24(2\tau^*)^2}e^{-L\pi h_{ab}/2\tau^*}\,,
\end{equation}
where $h_{ab}$ is the minimum (chiral) conformal weight of the fields that occur in the OPE $\Phi_{h_a}\times\Phi_{h_b}$. Note that $h_{aa}=0$ while $h_{ab}>0$ for $a\neq b$, i.e. the off-diagonal elements are  exponentially suppressed with the volume $L$ \cite{Cardy17}, and therefore the same is true for the level splitting generated by them.

We checked by explicit computation that in all cases considered in this work \eqref{matableading} gives the same eigenvectors lying in the subspaces $\mathcal{H}_{\text{QFT}}^{\pm}$ as the previous consideration, with the lower level always lying in the subspace $\mathcal{H}_{\text{QFT}}^{+}$. In fact, this result is easy to understand: replacing  $H$ by $H_{\text{CFT}}$ corresponds to setting $\lambda=0$, but once the appropriate eigenstates are determined to lie in the subspaces $\mathcal{H}_{\text{QFT}}^{\pm}$, switching on the perturbation cannot mix them. In addition, it cannot change their ordering in energy either due to the absence of level crossings along the RG flow generated by the perturbation.

\section{Chiral entanglement spectrum and entropy}
\label{sec:ent}
The Ansatz \eqref{Ansatz} immediately implies an analytic expression for the chiral reduced density matrix and the chiral entanglement entropy introduced in Section~\ref{sec:intro}. To include the possibility of degenerate solutions and linear combinations of physical boundary states, we consider a variational state 
\begin{equation}
|\Psi_\text{var}(L)\rangle=\frac{e^{-\tau^* H_\text{CFT}}}{\sqrt{\mathcal{N}_{\text{var}}}}\sum_{h}c_{a^* h}|\Phi_h\rangle\rangle\,,
\end{equation}
which is a linear combination of Ishibashi states as in Eq.~\eqref{CS}, but with \textit{a priori} arbitrary coefficients $c_{a^* h}$ (i.e. not necessarily a conformal boundary state). The normalization $\mathcal{N}_{\text{var}}$ is fixed by  $\langle\Psi_\text{var}(L)|\Psi_\text{var}(L)\rangle=1$ and is a suitable linear combination of Virasoro characters, cf. Eq.~\eqref{Virchar}. The overlaps of the variational state with the conformal basis are then
\begin{equation}
\label{overgs}
\langle\Psi_\text{var}(L)|h,N,k\rangle_l\otimes|h,N',k'\rangle_r=\delta_{NN'}\delta_{kk'}\frac{c_{a^* h}}{\sqrt{\mathcal{N}_\text{var}}}e^{-\frac{2\pi\tau^*}{L}(2h+2N-c/12)}\, ,
\end{equation}
and we used Eq.~\eqref{hamCFT} for the conformal Hamiltonian.  For a scalar perturbation the actual ground state of the QFT Hamiltonian~\eqref{gs_pert} has zero conformal spin  and therefore can be expanded in the conformal basis as
\begin{equation}
\label{gs}
|\Psi(L)\rangle=\sum_{h}\sum_{N}\sum_{k,k'}C^{hN}_{kk'}(L)|h,N,k\rangle_{l}\otimes|h,N,k'\rangle_{r};
\end{equation}
It is then  clear from Eq.~\eqref{overgs} that the variational state can only provide a valid approximation of the QFT ground state if off-diagonal elements of the matrices $C^{hN}$ are small and  diagonal ones satisfy Eq.~\eqref{overgs}. This is verified explicitly  in Section~\ref{sec:TCSA} using the TCSA.

The chiral reduced density matrix $\rho_r$ for the variational state defined in Eq.~\eqref{red} is then diagonal on the orthonormal basis $\{|h,N,k\rangle_r\}$:
\begin{equation}
\label{red_D}
\rho_r(L)=\frac{1}{\mathcal{N}_{\text{var}}}\sum_{h,N,k} |c_{a^* h}|^2 q^{h+N-c/24}|h,N,k\rangle_r{}_r\langle h,N,k|.
\end{equation}
where $q=e^{-8\pi\tau^*/L}$. 

The modular property \eqref{modular_transf} of the conformal characters allows an analytic evaluation of physical quantities in the limit $L\gg\tau$ as we now demonstrate for the chiral entanglement entropy. It is convenient to start with the \emph{$n$-th chiral R\'enyi entropy} for the variational state defined as
\begin{equation}
\label{reny1}
 \mathcal{S}^{n}_{\chi}=\frac{1}{1-n}\log\left[\text{Tr}\rho_{r}^n\right].
\end{equation}
Substituting the expression Eq.~\eqref{red_D} for the chiral reduced density matrix, \eqref{reny1} is expressed as \begin{equation}
\label{reny}
 \mathcal{S}^n_{\chi}=\frac{1}{1-n}\log\left[\frac{\sum\limits_{h}|c_{a^* h}|^{2n}\chi_{h}(q^n)}{\left(\sum\limits_{h}|c_{a^* h}|^{2}\chi_{h}(q)\right)^n}\right].
\end{equation}
The limit $L\gg\tau$ can be determined applying the modular transformation \eqref{modular_transf}, and retaining only the leading order contribution which consists of only that of  the conformal vacuum state:
\begin{equation}
\label{chireny}
\mathcal{S}^{n}_{\chi}\stackrel{L\gg\tau^*}{\simeq}\left(n+\frac{1}{n}\right)\frac{\pi c L}{48\tau^*}+\frac{1}{1-n}\log\left[\frac{\sum_{h}|c_{a^*h}|^{2n}S_{h}^{0}}{\left(\sum_{h}|c_{a^*h}|^{2}S_{h}^{0}\right)^n}\right].
\end{equation}
The chiral entanglement entropy of the variational ground state \eqref{Ansatz_f} is then obtained in the limit $n\rightarrow 1$:
\begin{equation}
\label{eq:LREE_final}
\mathcal{S}_{\chi}\equiv\lim_{n\rightarrow 1}\mathcal{S}^n_{\chi}\stackrel{L\gg\tau^*}{\simeq}\frac{\pi cL}{24\tau^*}-\gamma_{a^*},
\end{equation}
which shows an extensive dependence on the volume as anticipated in Eq.~\eqref{ent_sum}. The $O(1)$ term $\gamma_{a^*}$ in the large volume expansion is independent on the smearing parameter $\tau^*$ and reads
\begin{equation}
\label{gamma_top}
\gamma_{a^*}=\frac{\sum_{h}S_0^{h}|c_{a^*h}|^2\log\left(|c_{a^*h}|^2\right)}{\sum_{h}S_0^{h}|c_{a^*h}|^2}-\log\left(\sum_{h}S_0^{h}|c_{a^*h}|^2\right).
\end{equation}
Since the variational Ansatz~\eqref{Ansatz_f} is supposed to be exact in the limit $L\rightarrow\infty$, we conjecture that $\gamma_{a^*}$ in Eq.~\eqref{gamma_top} is a universal number characterising the ground state of a perturbed CFT in infinite volume. In Section~\ref{sec:TCSA} TCSA is used to verify numerically the validity of Eqs.~(\ref{eq:LREE_final},~\ref{gamma_top}) which  were derived assuming the variational Ansatz.

We close this section with a few comments and remarks. First, analogous expressions for the density matrix \eqref{red_D} and the universal term $\gamma_{a^*}$ in  Eq.~\eqref{gamma_top} contained in this section already appeared in~\cite{KL, DubailReadRezayi} (see also~\cite{DD}), in the context of topological states of matter~\cite{KP,LevinWen}. In~\cite{KL,DubailReadRezayi}, Eq.~\eqref{red_D} was in particular conjectured to explain the conformal nature of the entanglement spectrum~\cite{LH} in a $2+1$ d Quantum Hall fluid. Here we derived these results in the context of $1+1$ d massive QFT RG flows, as direct consequences of  the variational Ansatz \eqref{Ansatz_f}. Secondly, it is important that in our context the validity of these results is   numerically testable with TCSA.

Finally, notice that $\gamma_{a^*}$ is obtained in the opposite limit compared to the better-known Affleck--Ludwig boundary entropy~\cite{AL}. This observation allows establishing a partial analogy with recent studies~\cite{lr} concerning the real space entanglement spectrum in the ground state of a perturbed CFT. In~\cite{lr} the  entanglement Hamiltonian is the generator of translation around the annulus (i.e. in the periodic direction)  in Figure~\ref{fig:ent} with aspect ratio 
\begin{equation}
\frac{L}{2\tau}\simeq \frac{2\pi}{\log(\xi/a)}\,,
\end{equation} 
where  $\xi$  is the inverse of the mass gap and $a$ is an UV cut-off. The boundary conditions on the annulus are left free on one side and are fixed by a conformal boundary state on the other, although the boundary states are not explicitly determined in~\cite{lr}. The entanglement spectrum in real space displays universal features in the limit $\xi\gg a$ which corresponds to $L\ll\tau$, and the $O(1)$ term in the corresponding entanglement entropy is the Affleck--Ludwig boundary entropy, however due to the logarithmic dependence on the cut-off this result is not generally universal~\cite{CC}. 

In contrast, the chiral entanglement Hamiltonian following from \eqref{red_D} is the chiral part of the generator of translation along the annulus (i.e. in the open direction), with the same (linear combination of) conformal boundary states on both sides. As a result, the chiral entanglement spectrum displays universal features in the opposite  limit $L\gg\tau$, while the $O(1)$ term in the entropy is given by Eq.~\eqref{gamma_top} and is universal. Note that albeit the two settings are related by a modular transformation corresponding to crossing between the open and closed channels, the relevant boundary conditions are generally different.

\section{Examples}
\label{sec:exmp}
We now present applications of the formalism we discussed in the previous two sections. In the examples below the UV fixed point of the RG flow~\eqref{act} is a Virasoro minimal model~\cite{BPZ} with diagonal partition function on the torus. The central charge of the minimal model $M_p$ is 
\begin{equation}
c=1-\frac{6}{p(p+1)},\quad p=3,4,5,\dots
\end{equation}
while the primary fields $\Phi_{h_{r,s}}$ have scaling dimensions $2h_{r,s}$  with
\begin{equation}
h_{r,s}=\frac{[r(p+1)-sp]^2-1}{4p(p+1)},
\end{equation}
for $1\leq r\leq p-1$ and $1\leq s\leq p$. Taking into account the symmetry $h_{r,s}=h_{p-r,p+1-s}$, there are exactly $p(p-1)/2$ primary fields. The CFT  $M_p$ describes the $(p-1)$th Ising multicritical point, with discrete symmetry $\mathbb Z_2$~\cite{Zam_LG}, and explicit expressions for the  modular $S$ matrices can be found for instance in~\cite{Difra}:
\begin{eqnarray}
\chi_{r,s}(q)=&&\sum\limits_{r',s'}S_{rs}^{r's'}\chi_{r',s'}(\tilde q)\nonumber\\
&&S_{rs}^{r's'}=2\sqrt{\frac{2}{p(p+1)}}(-1)^{1+sr'+s'r}\sin\frac{\pi p r r'}{p+1}\sin\frac{\pi (p+1) s s'}{p}\,.
\end{eqnarray}

\subsection{Ising Field Theory} 
The simplest example is the perturbed Ising CFT with the fixed point $M_3$. The non-trivial relevant fields that could be considered as perturbations are  
\begin{itemize}
\item $\mathbb Z_2$ odd sector: spin field $\sigma=\Phi_{{2,2}}$ ($h_{2,2}=1/16$),
\item $\mathbb Z_2$ even sector: energy field $\varepsilon=\Phi_{{2,1}}$ ($h_{2,1}=1/2$)
\end{itemize}
The coupling to the spin field is denoted by $h$ and the one to the energy field by $t$, where one can always choose $h\geq 0$ since  $h<0$ can be obtained by $\mathbb Z_2$ symmetry $\sigma\rightarrow-\sigma$. 
As noticed in~\cite{Cardy17}, for $h=0$  the Ansatz does not reproduce the logarithmic singularity of the ground state energy correctly. However this contribution is a universal shift of all the energy eigenvalues and does not affect the state vectors themselves (cf. also Section~\ref{sec:TCSA}). The Cardy boundary states are given in terms of Ishibashi states as~\cite{Cardy89}
\begin{align}
\label{Isingbs1}
&|\widetilde{0}\rangle=\frac{1}{\sqrt{2}}|1\rangle\rangle+\frac{1}{\sqrt{2}}|\varepsilon\rangle\rangle+\frac{1}{2^{1/4}}|\sigma\rangle\rangle\\
\label{Isingbs2}
&|\widetilde{1/2}\rangle=\frac{1}{\sqrt{2}}|1\rangle\rangle+\frac{1}{\sqrt{2}}|\varepsilon\rangle\rangle-\frac{1}{2^{1/4}}|\sigma\rangle\rangle\\
\label{Isingbs3}
&|{\widetilde{1/16}}\rangle=|1\rangle\rangle-|\varepsilon\rangle\rangle,
\end{align} 
and have a simple interpretation in terms of spins in the lattice model~\cite{Cardy89}: the first two correspond to the boundary spins fixed in up/down position, while the third one describes the free boundary condition. The variational analysis leads to the following results~\cite{Cardy17}: 
\begin{itemize}
\item For $h>0$ and $t=0$ the variational solution is given by $|a^*\rangle=|\widetilde{1/2}\rangle$ in Eq.\eqref{Ansatz_f}. Applying  Eq.~\eqref{gamma_top}, the universal $O(1)$ term for the chiral entanglement entropy is 
\begin{equation}
\gamma_{\widetilde{1/2}}=-\frac{3}{4}\log(2)\,.
\end{equation} 
Including the thermal perturbation $t\neq 0$ this result is unaffected as long as $h/|t|^{15/8}$ is sufficiently large~\cite{Cardy17}. 
\item For $t>0$ and $h=0$, the ground state is best approximated by  $|a^*\rangle=|\widetilde{1/16}\rangle$, for which $\gamma_{\widetilde{1/16}}=0$. In the infinite volume limit the variational state coincides  with the ground state in the Neveu--Schwarz sector of a free Majorana fermion~\cite{Kon}.
\item The case $t<0$ and $h=0$ corresponds to the $\mathbb {Z}_2$ spontaneously broken phase and the variational Ansatz is degenerate: the two solutions are given by  $|a^*\rangle=|\widetilde{1/2}\rangle$ and $|b^*\rangle=|\widetilde{0}\rangle$. Their degeneracy is lifted in a finite volume and the corresponding eigenvectors can be identified following the method discussed in Subsection~\ref{subsec:degenerate}. The OPEs of the three conformal modules with the perturbing field are~\cite{BPZ}
\begin{align}
& 1\times\varepsilon=\varepsilon\,,\nonumber\\
 & \varepsilon\times\varepsilon=1\,,\nonumber\\
 & \sigma\times\varepsilon=\sigma\,.
\end{align}
Therefore the Hilbert space splits into the $\mathbb Z_2$-even (Neveu--Schwarz) sector $\mathcal{H}_{\text{QFT}}^{+}$ built upon the primaries $1$ and $\epsilon$, and the $\mathbb Z_2$-odd  (Ramond) sector $\mathcal{H}_{\text{QFT}}^{-}$ built upon the primary $\sigma$. 

The finite volume ground state is the unique linear combination of  the Cardy states $|\widetilde{0}\rangle$ and $|\widetilde{1/2}\rangle$ which belongs to $\mathcal{H}_{\text{QFT}}^+$:
\begin{equation}
\label{sym}
|\text{NS}\rangle\equiv
|\widetilde{0}\rangle+|\widetilde{1/2}\rangle=\sqrt{2}\left( |1\rangle\rangle+|\varepsilon\rangle\rangle \right)\,,
\end{equation}
for which the $O(1)$ term of the chiral entanglement entropy vanishes:
\begin{equation}
\gamma_{\text{NS}}=0\,.
\end{equation}
The antisymmetric linear combination  
\begin{equation}
\label{antisym}
|\text{R}\rangle\equiv
|\widetilde{0}\rangle-|\widetilde{1/2}\rangle=2^{3/4}|\sigma\rangle\rangle\,,
\end{equation}
belongs to $\mathcal{H}^{-}_{\text{QFT}}$  and corresponds to the first excited state which has a level spacing from the ground state 
$|\text{NS}\rangle$ which decays exponentially with the volume $L$. The smeared boundary state obtained form  Eq.~\eqref{antisym} has a different universal $O(1)$ constant in the chiral entanglement entropy:
\begin{equation}
\gamma_{\text{R}}=\log(\sqrt{2})\,.
\end{equation}

The same conclusions could be reached using the Kramers-Wannier duality~\cite{WuPotts} which maps $\varepsilon$ into $-\varepsilon$ and transforms the state $|\widetilde{1/16}\rangle$ into Eq.~\eqref{sym}, thereby mapping the variational solution for $t>0$ into the result for $t<0$.

Note that the two degenerate ground states in the ferromagnetic phase of the free Ising field theory are distinguished by a different $O(1)$ term in the chiral entanglement entropy, in contrast to their real space entanglement.
\end{itemize}

\subsection{Tricritical Ising Field Theory}

The minimal model $M_4$ describes the universality class of the tricritical Ising model~\cite{DFQ, chri_mus} and has a global $\mathbb{Z}_2$ symmetry with six primary fields organized under the $\mathbb Z_2$ symmetry  as~\cite{Zam_LG}
\begin{itemize}
\item $\mathbb{Z}_2$ odd sector: $\sigma=\Phi_{h_{2,2}}$ ($h_{2,2}=3/80$),  $\sigma'=\Phi_{h_{2,1}}$ ($h_{2,1}=7/16$)
\item $\mathbb{Z}_2$ even sector: $\varepsilon=\Phi_{h_{3,3}}$ ($h_{3,3}=1/10$), $\varepsilon'=\Phi_{h_{3,2}}$ ($h_{3,2}=3/5$), $\varepsilon''=\Phi_{h_{3,2}}$ ($h_{3,2}=3/2$)
\end{itemize}
The couplings to the $\mathbb Z_2$ odd fields $\sigma$ and $\sigma'$ are denoted by $h$ and  $h'$, while couplings to the $\mathbb Z_2$ even fields $\varepsilon$ and $\varepsilon'$ are denoted by $t$ and $t'$, respectively (note that the field $\varepsilon''$ is irrelevant).  The microscopic interpretation of the different physical boundary states was given in~\cite{Affleck_tri} in the context of the dilute Ising model. 

For the sake of brevity we only consider a perturbations by a single field; the generalization to more than one perturbing fields is straightforward. The variational analysis leads to the following results (by convention all couplings not explicitly specified are assumed to be zero): 
\begin{itemize}
\item $h>0$: the variational solution is obtained by selecting $|a^*\rangle=|\widetilde{3/2}\rangle$ in Eq.~\eqref{Ansatz_f} which corresponds to a boundary condition with all spins fixed.
\item $t>0$: the unique solution is $|a^*\rangle=|\widetilde{7/16}\rangle$, corresponding to a $\mathbb Z_2$ symmetric ground state. 
\item $t<0$: analogously to the Ising case there are two degenerate solutions of~\eqref{varE}  corresponding to the Cardy states 
$|\widetilde{0}\rangle$ and $|\widetilde{3/2}\rangle$. The two sectors $\mathcal{H}_{\text{QFT}}^\pm$ correspond again to even/odd primaries. In a finite volume the true ground state is in the even sector and is given by the smeared boundary state \eqref{Ansatz_f} obtained from
\begin{equation}
|\text{tNS}\rangle\equiv|\widetilde{0}\rangle+|\widetilde{3/2}\rangle=\mathcal{N}\bigl(|1\rangle\rangle+|\varepsilon''\rangle\rangle+\sqrt{\varphi}|\varepsilon'\rangle\rangle+\sqrt{\varphi}|\varepsilon\rangle\rangle\bigr),
\label{tNS}\end{equation}
where $\varphi=(1+\sqrt{5})/2$ is the golden ratio and $\mathcal{N}$ a numerical coefficient. The lowest energy state in the odd sector is given by
\begin{equation}
|\text{tR}\rangle\equiv|\widetilde{0}\rangle-|\widetilde{3/2}\rangle=\mathcal{N}'\bigl(|\sigma'\rangle\rangle+\sqrt{\varphi}|\sigma\rangle\rangle\bigr),
\label{tR}\end{equation}
where $\mathcal{N}'$  is another numerical coefficient. The two states are degenerate in the thermodynamic limit and their splitting decays exponentially with the volume.

Note that the even resp. odd combinations only contain fields in the Neveu--Schwarz  resp. Ramond sectors of the super-Virasoro algebra of the minimal model $M_4$~\cite{DFQ}, respectively (indeed this explains our choice of labelling for them).  The difference of the chiral entanglement entropies is given by
\begin{equation}
\label{dif_gamma}
\gamma_\text{tR}-\gamma_\text{tNS}=\log(\sqrt{2}),
\end{equation}
which coincides with the analogous expression $\gamma_\text{R}-\gamma_\text{NS}$ in the Ising case.

\item $h'>0$: again there are two degenerate solutions to the variational equations corresponding to the Cardy states $|\widetilde{0}\rangle$ and $|\widetilde{3/5}\rangle$ with associated primaries $\Phi_{0}=1$ and $\Phi_{3/5}=\varepsilon'$. The way the degeneracy is lifted in a finite volume can be obtained by examining the relevant OPEs~\cite{Difra}:
\begin{equation}
 \begin{array}{l}
  1\times \sigma'=\sigma'\\
  \varepsilon''\times\sigma'=\sigma'\\
  \sigma'\times\sigma'=1+\varepsilon''
 \end{array},\qquad
 \begin{array}{l}
  \sigma\times \sigma'=\varepsilon+\varepsilon'\\
  \varepsilon\times\sigma'=\sigma\\
  \varepsilon'\times\sigma'=\sigma
 \end{array};
\end{equation}
from which we can infer that sector $\mathcal{H}_{\text{QFT}}^{+}$ contains representations built from the primaries $\{1,\varepsilon'', \sigma'\}$, while $\mathcal{H}_{\text{QFT}}^{-}$ consists of $\{\sigma,\varepsilon, \varepsilon'\}$. The unique linear combination of degenerate Cardy states belonging to $\mathcal{H}_{\text{QFT}}^{+}$ is 
\begin{equation}
 |\text{GS}_+\rangle=|\widetilde{0}\rangle+\varphi|\widetilde{3/5}\rangle\,,
\end{equation}
while the one giving the lowest energy state in $\mathcal{H}_{\text{QFT}}^-$ is 
\begin{equation}
 |\text{GS}_-\rangle=-\varphi|\widetilde{0}\rangle+|\widetilde{3/5}\rangle\,,
\end{equation}
or in terms of Ishibashi states as
\begin{align}
\label{dc1}
 &|\text{GS}_+\rangle=\mathcal{N}\left[\frac{1}{\sqrt{2}}|1\rangle\rangle+\frac{1}{\sqrt{2}}|\varepsilon''\rangle\rangle+\frac{1}{2^{1/4}}|\sigma'\rangle\rangle\right]\\
 \label{dc2}
 &|\text{GS}_-\rangle=\mathcal{N}'\left[\frac{1}{\sqrt{2}}|\varepsilon\rangle\rangle+\frac{1}{\sqrt{2}}|\varepsilon'\rangle\rangle+\frac{1}{2^{1/4}}|\sigma\rangle\rangle\right]\,.
\end{align}
Moreover, using Eq.~\eqref{gamma_top}, the difference between their chiral entanglement entropy is given by
\begin{equation}
\gamma_{\text{GS}_-}-\gamma_{\text{GS}_+}=-\log(\varphi)\,.
\end{equation}
Comparing to Eq.~\eqref{Isingbs1} one can note that $|\text{GS}_+\rangle$ and $|\text{GS}_-\rangle$  are effectively two Ising-type boundary states; in particular, the fields $\{1, \sigma', \varepsilon''\}$ realize an OPE isomorphic to that of the Ising model. This observation confirms  that the tricritical Ising model perturbed by $\sigma'$ describes a phase separation between two magnetized Ising pure phases that are not distinguished by 
$\mathbb Z_2$ symmetry alone~\cite{Cardy_book}.

From the QFT point of view this conclusion is quite remarkable since it implies a double degeneracy of the ground state of different nature respect to the most familiar one observed in Ising field theory, which is potentially interesting to the study of topological phases of matter~\cite{FTLTKWF} (cf. also Section~\ref{sec:conc}).

\item $h'<0$: the degenerate Cardy states are $|\widetilde{3/2}\rangle$ and $|\widetilde{1/10}\rangle$.  Repeating the previous analysis gives the same states (\ref{dc1}) and (\ref{dc2}) but with a negative coefficient in front of the $\mathbb Z_2$-odd terms $|\sigma\rangle\rangle$ and $|\sigma'\rangle\rangle$. Similarly to the case $h'>0$, note again the parallel with the Ising case, but this time the analogous state is given by \eqref{Isingbs2}.

\end{itemize}

So far we only gave the difference of the chiral entanglement entropy between degenerate ground states where applicable. To complete this information, the values of the $O(1)$ universal contributions for the true ground states themselves are summarised in Table~\ref{Tab_tri}.
\begin{table}
\centering
\begin{tabular}{|c|c|l|}
\hline 
Perturbation &  Coupling & ~~~~~$\gamma$ for the true ground state\\ 
\hline 
$\sigma$ & $h>0$  & $\gamma_{\widetilde{3/2}}=4[s_1^2\log(s_1)+s_2^2\log(s_2)]+\frac{1}{4}\log(2)$ \\ 
\hline 
$\varepsilon$ & $t>0$  & $\gamma_{\widetilde{7/16}}=4[s_1^2\log(s_1)+s_2^2\log(s_2)]+\log(2)$  \\ 
\hline 
$\varepsilon$ & $t<0$  & $\gamma_\text{tNS}=\frac{s_1\varphi\log(\varphi)}{s_2+s_1\varphi}-\log[2(s_2+s_1\varphi)]$ \\ 
\hline 
$\sigma'$ & $h'>0$  &  $\gamma_{\text{GS}_+}=-\log(s_2)-\frac{7}{4}\log(2)$\\ 
\hline 
\end{tabular} 
\caption{$O(1)$ universal contributions to the chiral entanglement entropy for the true ground state in the perturbed tricritical Ising model, with the notations $s_1=\sin(2\pi/5)/\sqrt{5}$ and $s_2=\sin(4\pi/5)/\sqrt{5}$ (note that $s_1/s_2=\varphi$).}
 \label{Tab_tri}
\end{table}

\section{Numerical Results from TCSA}
\label{sec:TCSA}

Now we turn to a numerical computation of the chiral entanglement using the Truncated Conformal Space Approach (TCSA), which is a variational method originally introduced to construct the approximate low energy spectrum of field theories in finite volume in~\cite{YurZam}. Since then it was applied for various models and its scope extended to extracting a range of quantities as well as determining phase diagrams and simulating non-equilibrium time evolution; for a recent review we refer the interested reader to~\cite{JKLRT}. The ingredients necessary for the present computations are briefly reviewed in Appendices~\ref{appA}~and~\ref{appB}.
 
The TCSA consists of considering the Hamiltonian~\eqref{gs_pert} in finite volume with appropriate boundary conditions, which are chosen to be periodic for the subsequent calculations. The finite volume spectrum is discrete and introducing an upper energy cut-off reduces the Hilbert space to a finite-dimensional subspace, in which all states and operators are represented as finite-dimensional vectors and matrices, respectively. Units are chosen by expressing the coupling in terms of the mass gap $m$ (i.e. the mass of the lightest excitation), and the volume is parameterized by the dimensionless combination $mL$, while energies are measured in units of $m$. The determination of the spectrum is reduced to the diagonalization of the finite truncated Hamiltonian matrix, with the components of eigenstates numerically known in the conformal basis. As a result, the reduced density matrix and the chiral entanglement entropy can be computed in a straightforward way. The results obtained from TCSA depend on the cut-off: the more relevant the perturbing operator the faster the convergence with the cut-off. The convergence can be improved further by eliminating the leading cut-off dependence using a renormalization group approach, explained in Appendix~\ref{appB} for the case of the chiral entanglement entropy.

\subsection{Verifying the Ansatz: energy densities and overlaps}

Before considering the chiral entanglement entropy, we first test Cardy's variational Ansatz. The simplest test is to compare the predicted ground state energy density $\mathcal{E}_{predicted}$ obtained by substituting the variational solution into \eqref{varE} to exact results coming from integrability~\cite{Fat94}, or from the one determined numerically from TCSA when the perturbation is non-integrable. The results of this comparison are summarized in Table~\ref{tab:taus}. Note that for the thermal perturbation of the Ising model the energy density predicted by the Ansatz is not meaningful since there is a logarithmic divergence with the UV cut-off. The results demonstrate that the Ansatz predicts the energy density for strongly relevant perturbations quite precisely, however its accuracy gets progressively worse when the weight of the perturbation increases.

\begin{table}
\centering
\begin{tabular}{|c|c|c|c|c|}
\hline 
Model & $\kappa$ & $m \tau^*$ & $\mathcal{E}_{predicted}/m^2$ & $\mathcal{E}_{theory}/m^2$ \\ 
\hline 
IFT+$\sigma$ & $0.06203236$  & $1.9970027$ & $-0.0615441$ & $-0.0617286$ \\ 
\hline 
IFT+$\varepsilon$ & $\pm 0.159155$ & $0.261799$ & -- & -- \\
\hline
TIFT+$\sigma$ $^*$ & 0.1 & $2.0844592$ & $-0.135319$ & $-0.135496$ \\
\hline
TIFT+$\varepsilon$ & $\pm 0.0928344$ &  $1.484332$ & $-0.0935744$ & $-0.0942097$ \\
\hline
TIFT+$\sigma'$ & $\pm 0.166252$ & $0.370412$ & $-0.214659$ & $-0.415603$ \\
\hline
\end{tabular} 
\caption{Summary of the results from Cardy's Ansatz. The second column shows the value of $\kappa$ (cf. \eqref{massgap}), which for integrable models was chosen so that the mass unit $m$ equals the mass gap of the model. The model labeled by star is non-integrable, therefore the choice of $\kappa$ is arbitrary resulting in some fixed scale $m$ which is not equal to the mass gap, and for this case the theoretical prediction is replaced by the numerical value obtained from the  TCSA ground state by extrapolation using cut-off levels from $5$ to $12$.}
\label{tab:taus}
\end{table}

\begin{figure}
\centering
\includegraphics{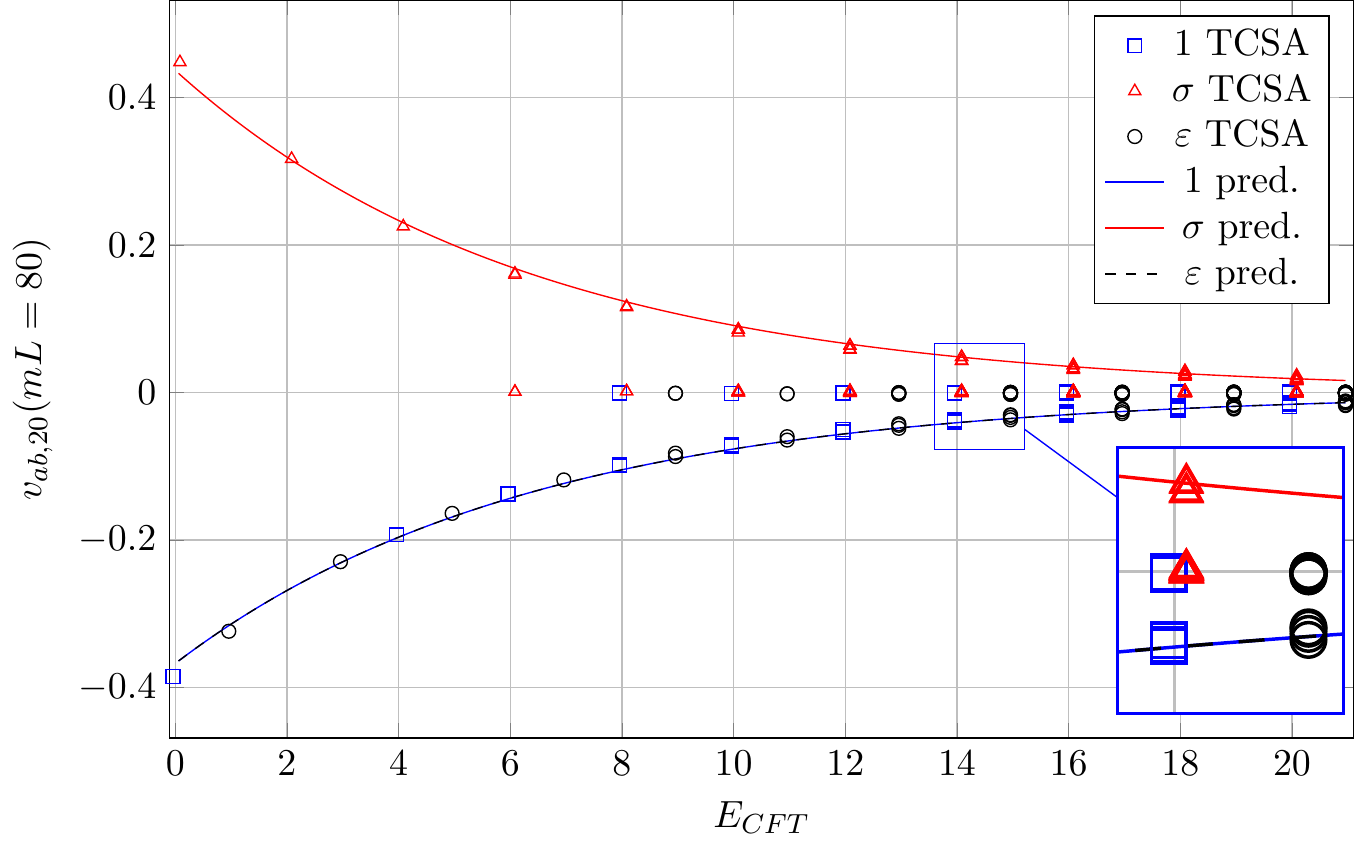}
\caption{Coefficients of the ground state eigenvector as a function of the conformal energy (eigenvalue of $L_0+\bar{L}_0-c/12$) in Ising field theory perturbed by the spin field $\sigma$ ($t=0,\ h>0$) at the dimensionless volume $mL=80$. Discrete dots are TCSA data at level $20$ with $28624$ states, while the lines show the prediction of~\eqref{overgs} with $\tau_*$ given in  Table~\ref{tab:taus} (with an overall sign difference which is due to the choice of the numerics). Different colours correspond to different modules: blue--$1$, red--$\sigma$ and black--$\varepsilon$; the point lying on (or close to) the horizontal axis correspond to non-diagonal contributions. Inset: zooming on the conformal energy region $\approx14-15$ shows the deviations from the diagonal form of the Ansatz discussed in the text.}
\label{fig:E8gsvec}
\end{figure}

\begin{figure}
\centering
\includegraphics{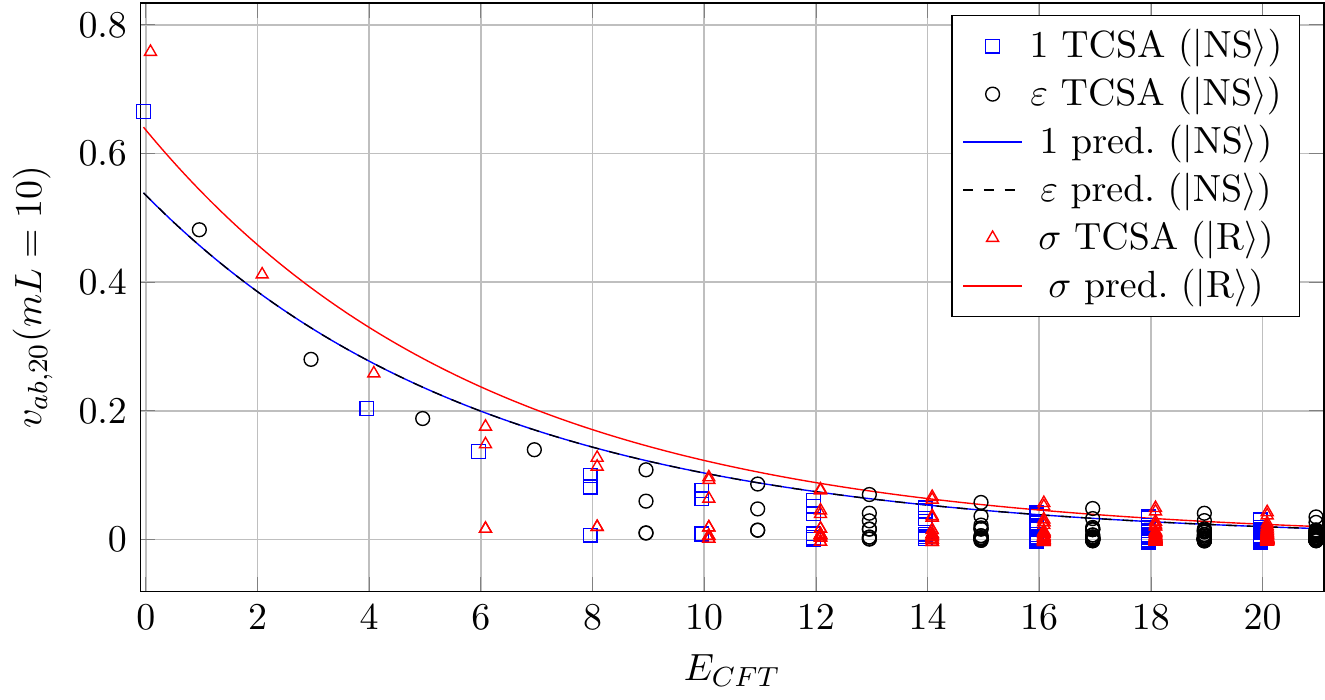}
\caption{Coefficients of the ground state vector as a function of the conformal energy (eigenvalue of $L_0+\bar{L}_0-c/12$) of the CFT state in the thermal perturbation of the Ising fixed point at $mL=10$. Dots are from TCSA at level $20$ with $28624$ states, the lines are predicted by using~\eqref{overgs} the smearing from Cardy's Ansatz (see Table~\ref{tab:taus}) and $c_{a^*h}$ corresponding to $|\text{NS}\rangle$ and $|\text{R}\rangle$. It is clear the overlaps are not in full quantitative agreement. The reason is that the conformal weight of the perturbation is high: the energy levels are logarithmically divergent, and the cut-off effects are relatively high. However, it turns out that the cut-off extrapolated chiral entanglement entropy calculated from the overlaps agrees very well with the theoretical prediction  (cf. Section~\ref{sec:TCSA}).}
\label{fig:m34_1o2_gsvec}
\end{figure}

\begin{figure}
\centering
\includegraphics{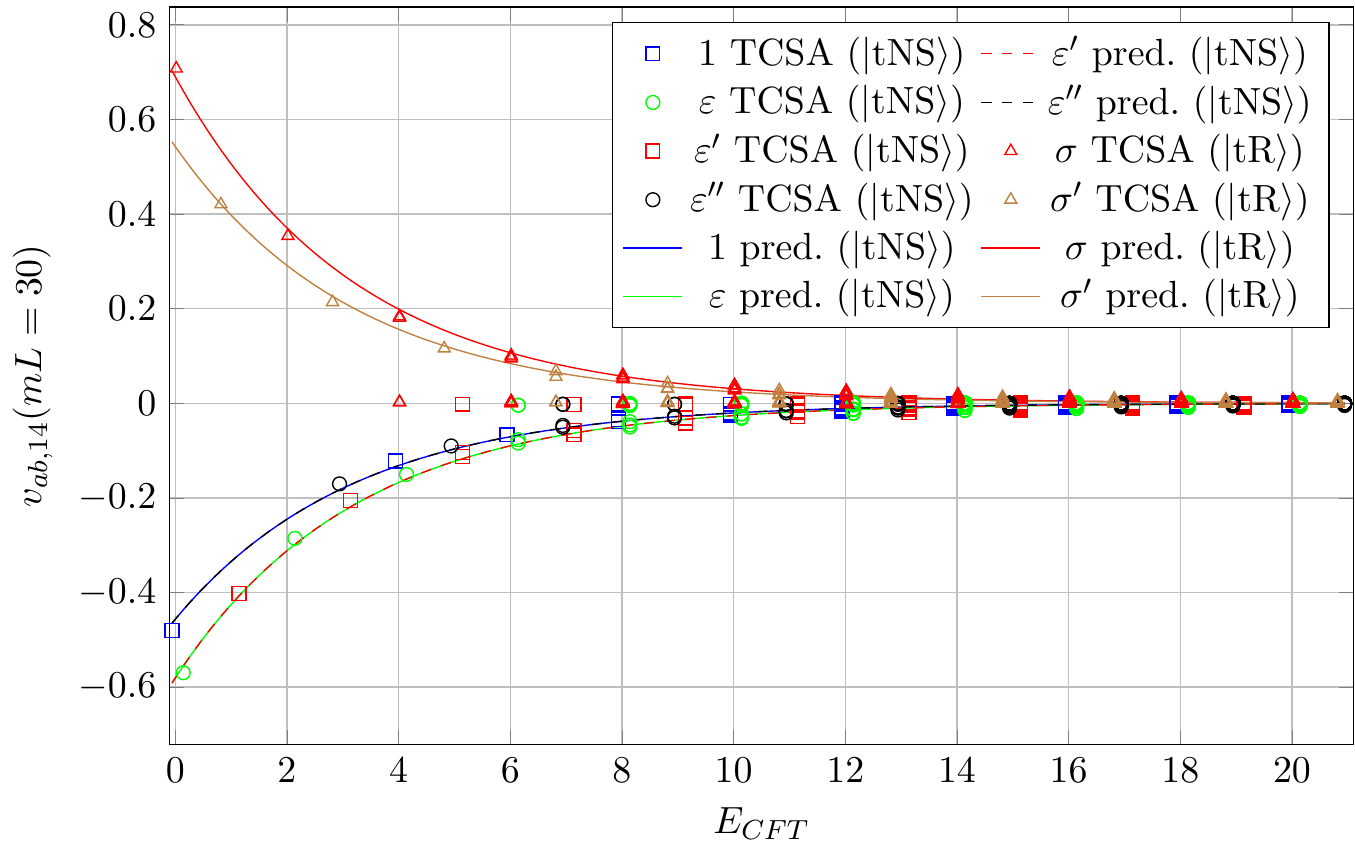}
\caption{Coefficients of the ground state vector as a function of the conformal energy (eigenvalue of $L_0+\bar{L}_0-c/12$) of the CFT state in the $t<0$ perturbation of the tricritical Ising fixed point at $mL=30$. Dots are from TCSA at level $14$ with $22559$ states in the NS sector and $18751$ in the R sector. The lines are predicted by using~\eqref{overgs} the smearing from Cardy's Ansatz (see Table~\ref{tab:taus}) and $c_{a^*h}$ corresponding to $|\text{tNS}\rangle$ and $|\text{tR}\rangle$.}
\label{fig:m45_1o10_gsvec}
\end{figure}

\begin{figure}
\centering
\includegraphics{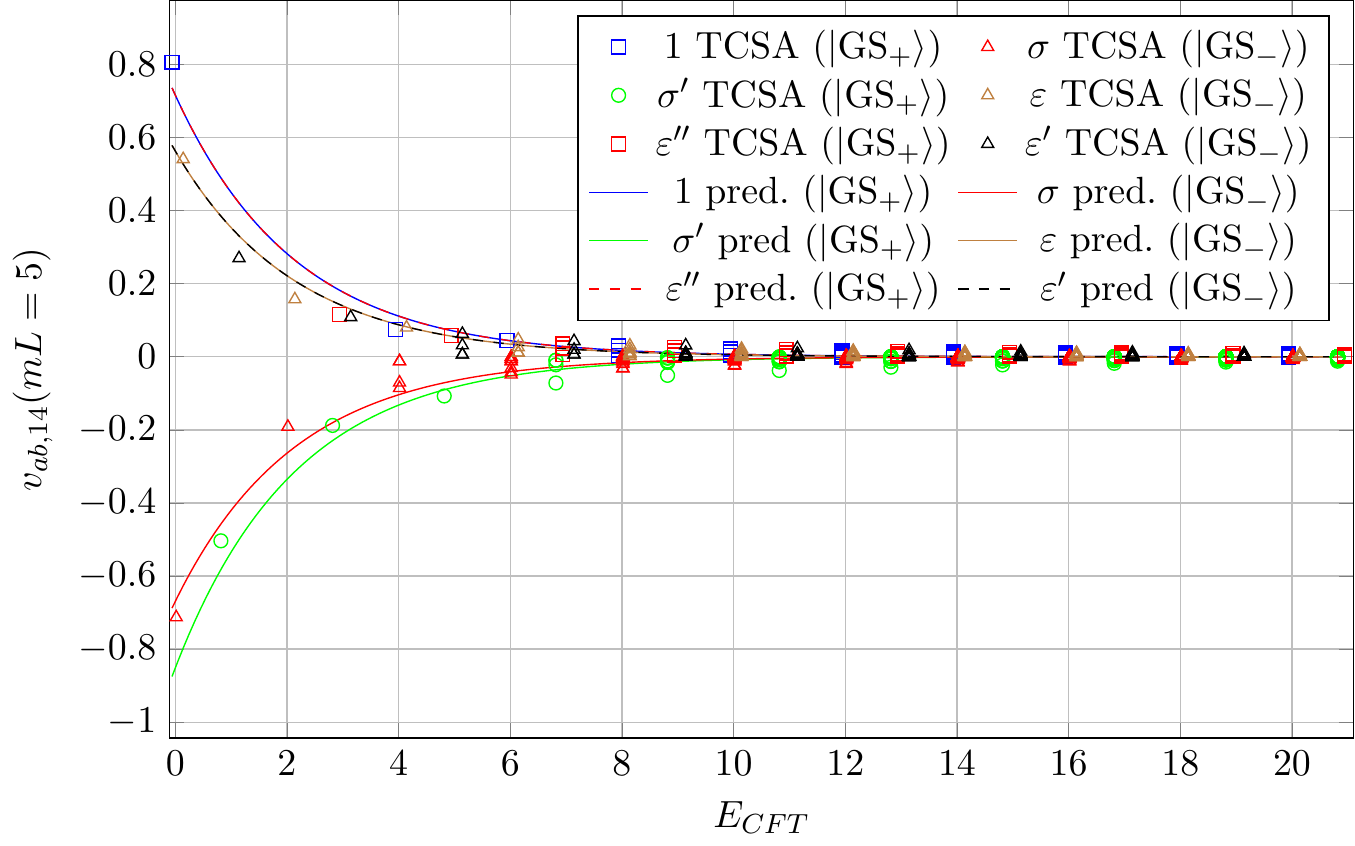}
\caption{Coefficients of the ground state vector as a function of the conformal energy (eigenvalue of $L_0+\bar{L}_0-c/12$) of the CFT state in the $h'>0$ perturbation of the tricritical Ising fixed point at $mL=5$. Dots are from TCSA at level $14$ with $13373$ states in the sector $\{1,\sigma',\varepsilon''\}$ and $27937$ in $\{\sigma,\varepsilon,\varepsilon'\}$. The lines are predicted by using~\eqref{overgs} the smearing from Cardy's Ansatz (see Table~\ref{tab:taus}) and $c_{a^*h}$ corresponding to $|\text{GS}_+\rangle$ and $|\text{GS}_-\rangle$.}
\label{fig:m45_7o16_gsvec}
\end{figure}

One can also compare the overlaps $v_{ab,\Lambda}(L)=\langle\Psi_{\Lambda}(L)|| a\rangle_l\otimes |b\rangle_r$ calculated from TCSA to the prediction from~\eqref{overgs} using the $c_{a^*h}$ coefficients obtained by the variational method. The results are demonstrated in Figures~\ref{fig:E8gsvec}, \ref{fig:m34_1o2_gsvec}, \ref{fig:m45_1o10_gsvec} and~\ref{fig:m45_7o16_gsvec}. The exponential fall-off of the coefficients with the conformal level is clearly visible and agrees with the predictions of the Ansatz. However, it is also clear that there are some discrepancies since the boundary states only have diagonal components in an orthonormal basis according to the structure of Ishibashi states \eqref{CS}, which should be exactly the same for components with a given highest weight $h$ and level $N$. The TCSA shows that this is violated: there are non-zero non-diagonal components, and the diagonal ones have slightly different values as well, in some cases they clearly deviate from the predicted exponential curve. These deviations indicate that the Ansatz is not perfect and is only strictly valid in the limit $L\rightarrow\infty$, and for a more precise description in finite volume it is necessary to include more irrelevant operators around the boundary fixed point as  pointed out in~\cite{Cardy17}. However as we show in Appendix~\ref{appB} (cf. Figure~\ref{fig:maverick}) both the (relative) difference between the diagonal coefficients and the non-diagonal coefficients converge to zero with increasing volume. 

The  variational Ansatz ~\eqref{Cardy_var} and its extension in presence of degeneracy are quite poor for Ising field theory perturbed by the energy operator; see Figure \ref{fig:m34_1o2_gsvec}. This is not surprising since it was already pointed out in \cite{Cardy17} that the Ansatz cannot account for the logarithmic contribution to the ground state energy either. Exact calcuation performed in~\cite{Kon} shows that the ratio of overlaps  between the true  QFT ground state and two states in the CFT basis can converge to the IR value (obtained for $mL\rightarrow\infty$) with a power law behaviour, rather than exponentially  as predicted instead by~\eqref{overgs}. It is  worth noticing, however, that the agreement obtained  for the chiral entanglement entropy remains  remarkable; see Figure~\ref{fig:LREE_ThermalIF}.

\subsection{Chiral entanglement entropy}
Now we turn to the predictions for the chiral entanglement entropy, which for large enough volume can be written in the form  \eqref{ent_sum}:
\begin{equation}
\label{ent_sum1}
\mathcal{S}_{\chi}(L)\sim \mathcal{B}L-\gamma+O(e^{-L}).
\end{equation}
The prediction therefore consists of two quantities: the linear slope $\mathcal{B}$, 
which according to \eqref{eq:LREE_final} can be obtained as 
\begin{equation}
\mathcal{B}=\frac{\pi c}{24\tau^*}
\end{equation}
from the UV central charge $c$ and the variational solution for $\tau^*$ listed in Table \ref{tab:taus}, and the $O(1)$ contribution $\gamma$
which was obtained in Section~\ref{sec:exmp}. 

\subsubsection{Ising Field Theory} 

In the case of the Ising field theory we studied numerically three different perturbations of the fixed point CFT.

\begin{figure}
 \centering
 \includegraphics{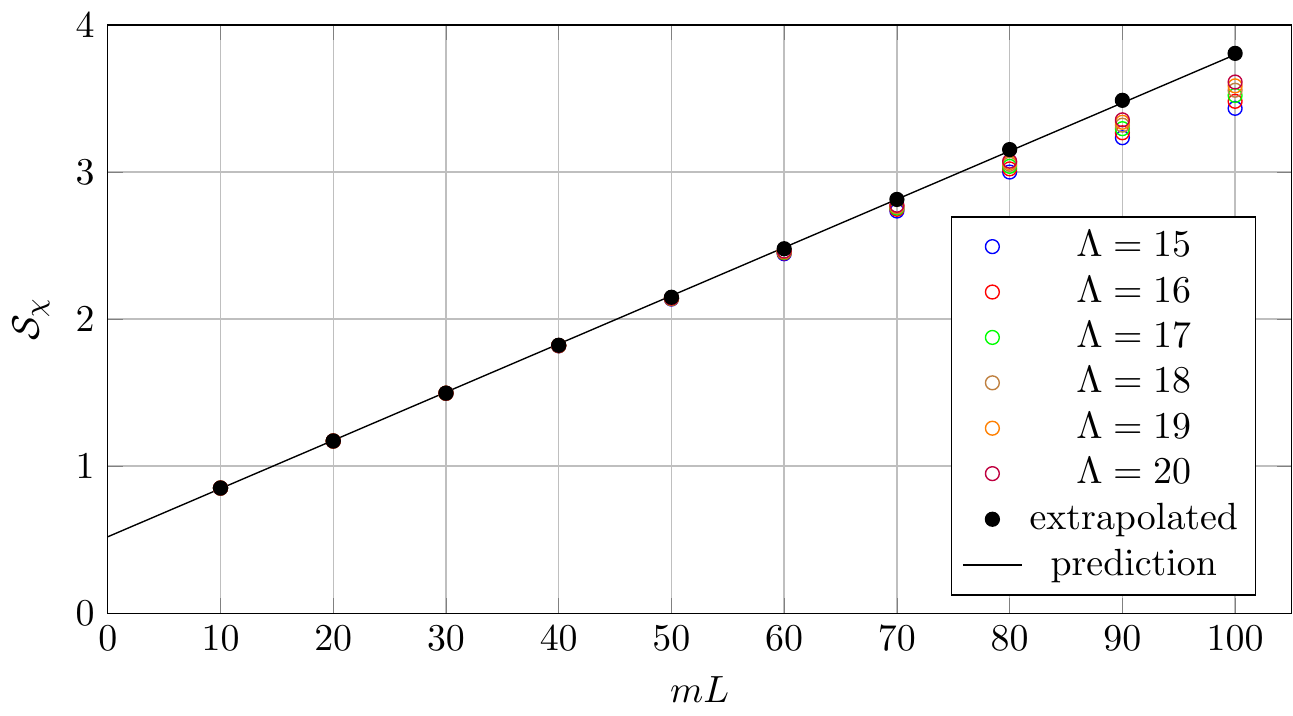}
 \caption{Chiral entanglement entropy in the magnetic perturbation of the Ising model. TCSA (circles/dots) against eq.~\eqref{eq:LREE_final} with the corresponding value of $\tau^*$ (see Table~\ref{tab:taus}). TCSA data are shown at different cut-offs and also after extrapolation.}
 \label{fig:LREE_E8}
\end{figure}

\begin{figure}
 \centering
  \begin{subfigure}[Paramagnetic phase, ground state corresponding to $|\widetilde{1/16}\rangle$.]{\includegraphics{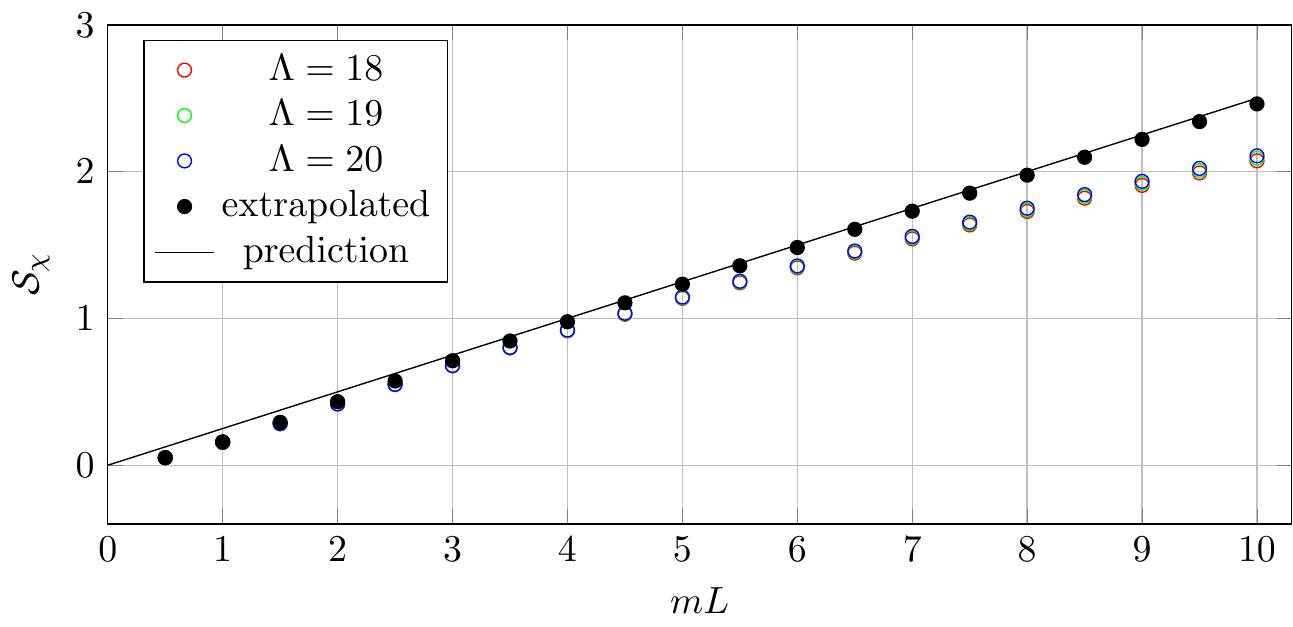}}
  \end{subfigure} 
  \begin{subfigure}[ Ferromagnetic phase, Neveu--Schwarz and Ramond ground states.]{\includegraphics{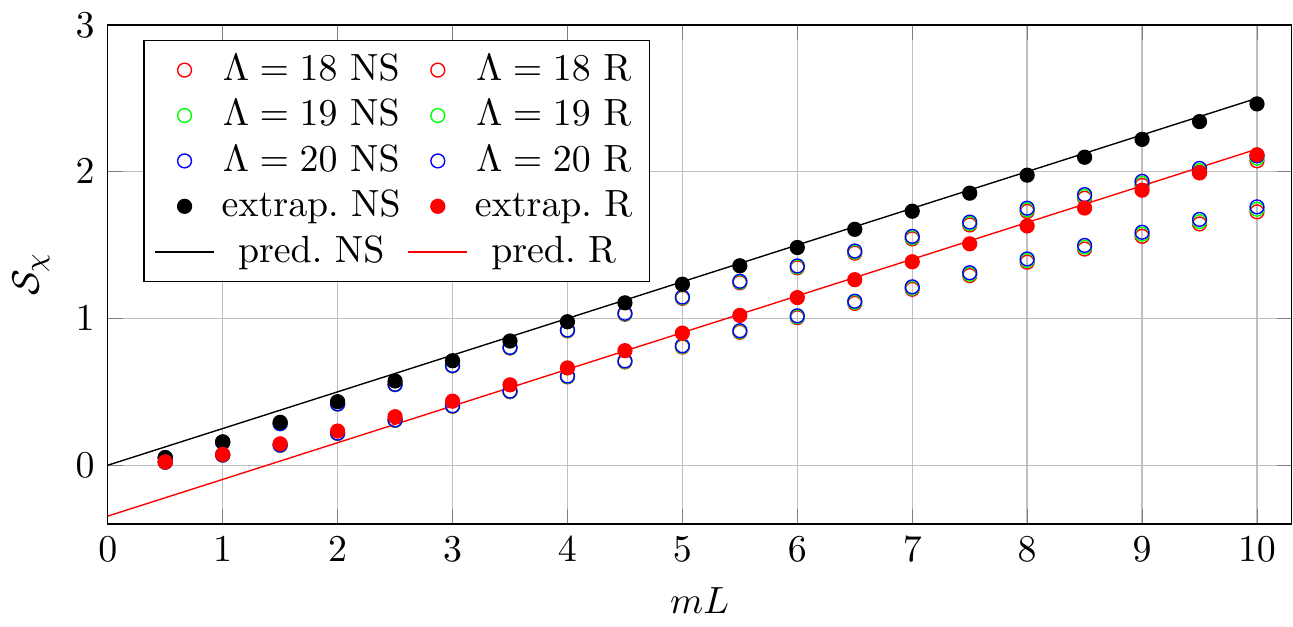}
}
 \end{subfigure}
 \caption{Chiral entanglement entropy in the thermal Ising field theory.}
 \label{fig:LREE_ThermalIF}
\end{figure}

\begin{table}
\centering
\begin{tabular}{|c|c|c|}
\hline 
Volume &  cut-off level $20$ & Extrapolated\\ 
\hline 
 $6$ & $0.33993$ & $0.34105$ \\ 
\hline 
$7$ & $0.34363$  & $0.34441$  \\ 
\hline 
$8$ & $0.34549$  & $0.34577$ \\ 
\hline 
$9$ & $0.34663$  &  $0.34628$\\ 
\hline
$10$ & $0.34753$  &  $0.34641$\\ 
\hline
\end{tabular} 
\caption{Difference between the chiral entanglement entropy of the ground state in the Neveu--Schwarz and the Ramond sector, showing both the bare TCSA value obtained for the highest available cut-off and the one obtained after cut-off extrapolation. The predicted value is $\log \sqrt{2}\approx0.346547$.}
 \label{tab:logsqrt2}
\end{table}

\begin{itemize}

\item $t=0,\ h>0$ (magnetic perturbation): as already demonstrated in Figure \ref{fig:E8gsvec} the variational state resulting from Cardy's Ansatz provides an excellent approximation to the ground state. The chiral entanglement entropy as a function of the volume is showed in Figure~\ref{fig:LREE_E8}, and it is clear that the predictions are in excellent agreement with the extrapolated TCSA value.

\item $t\neq 0,\ h=0$ (thermal perturbation): the positive/negative signs of the coupling correspond to the paramagnetic/ferromagnetic phases, for which the chiral entanglement entropy as a function of the volume is shown in Figure~\ref{fig:LREE_ThermalIF}. In the paramagnetic phase ($t>0$) the unique ground state $|\widetilde{1/16}\rangle$ is in the Neveu--Schwarz sector, while in the ferromagnetic case there are two degenerate ground states in large but finite volume, namely $|\text{NS} \rangle$ in the Neveu--Schwarz and $|\text{R}\rangle$ in the Ramond sector. In all cases the predictions from Cardy's Ansatz combined with eq.~\eqref{eq:LREE_final} are in agreement with the extrapolated TCSA data. In Table~\ref{tab:logsqrt2} we present the  difference between the chiral entanglement in the two ground states in different volumes, which shows that it converges to the predicted value with increasing volume.
\end{itemize}

\subsubsection{Tricritical Ising Field Theory}    

\begin{figure}
 \centering
 \begin{subfigure}[ $t>0$]{\includegraphics{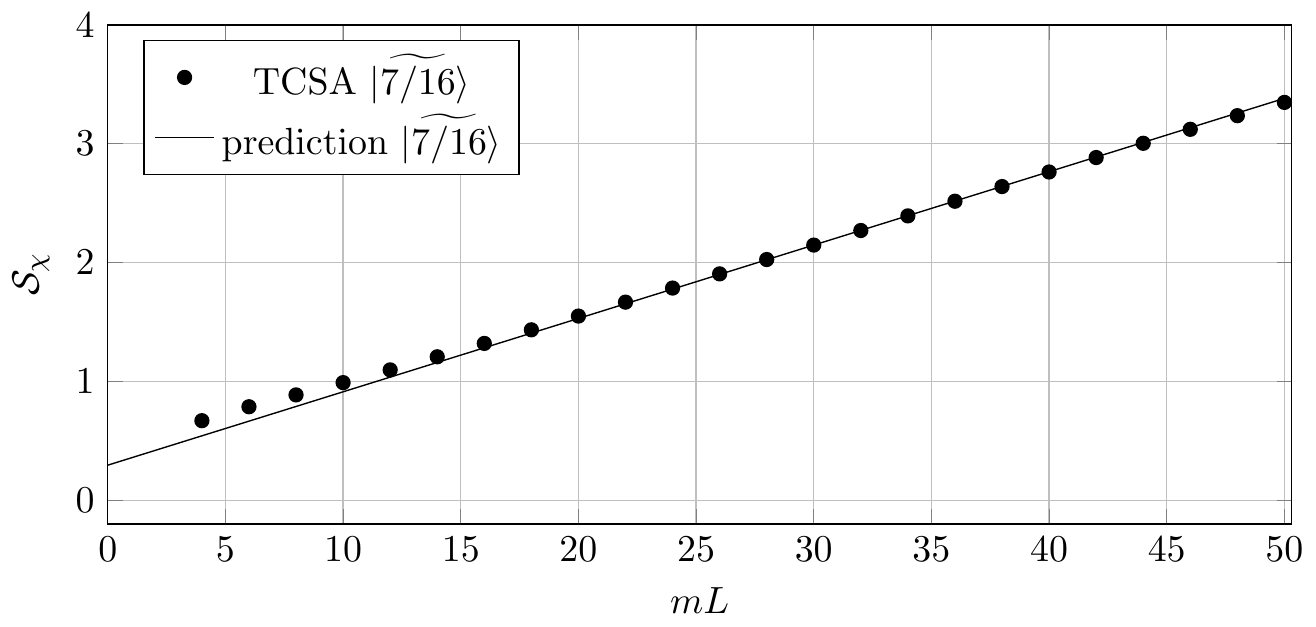}}
 \end{subfigure}
 \begin{subfigure}[ $t<0$]{\includegraphics{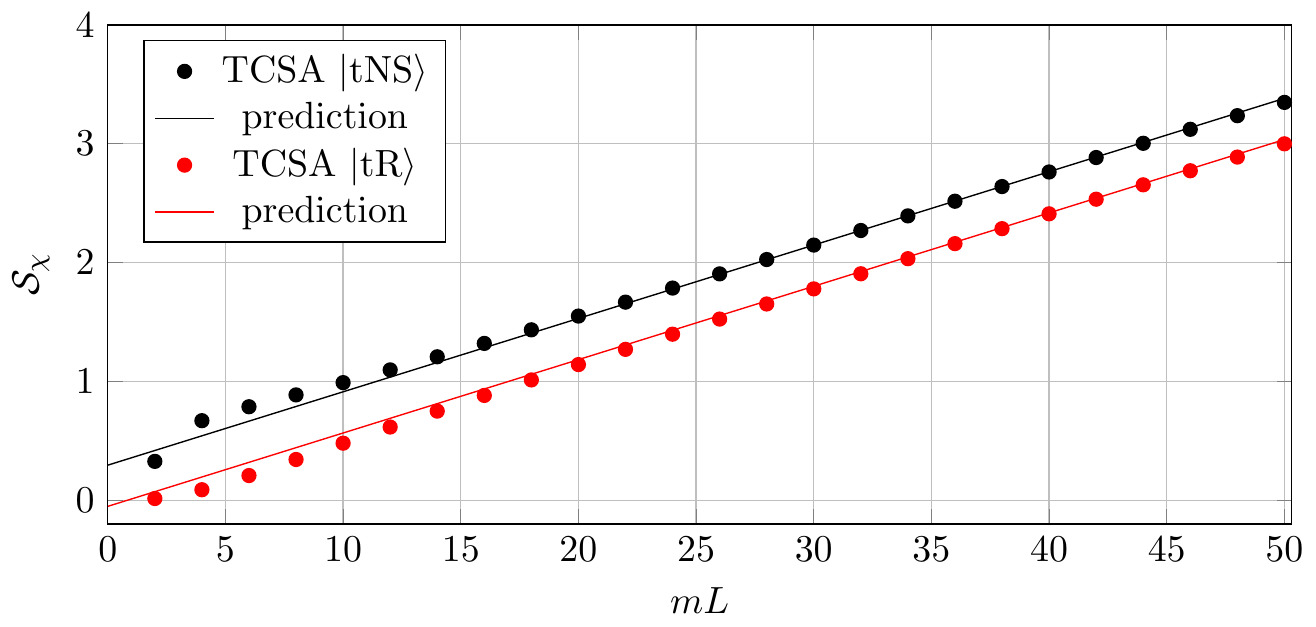}}
 \end{subfigure}
 \caption{Chiral entanglement entropy in TIFT+$\varepsilon$. TCSA data are extrapolated to infite cut-off.}
 \label{fig:LREE_m45_1o10}
\end{figure}

\begin{figure}
 \centering
 \includegraphics{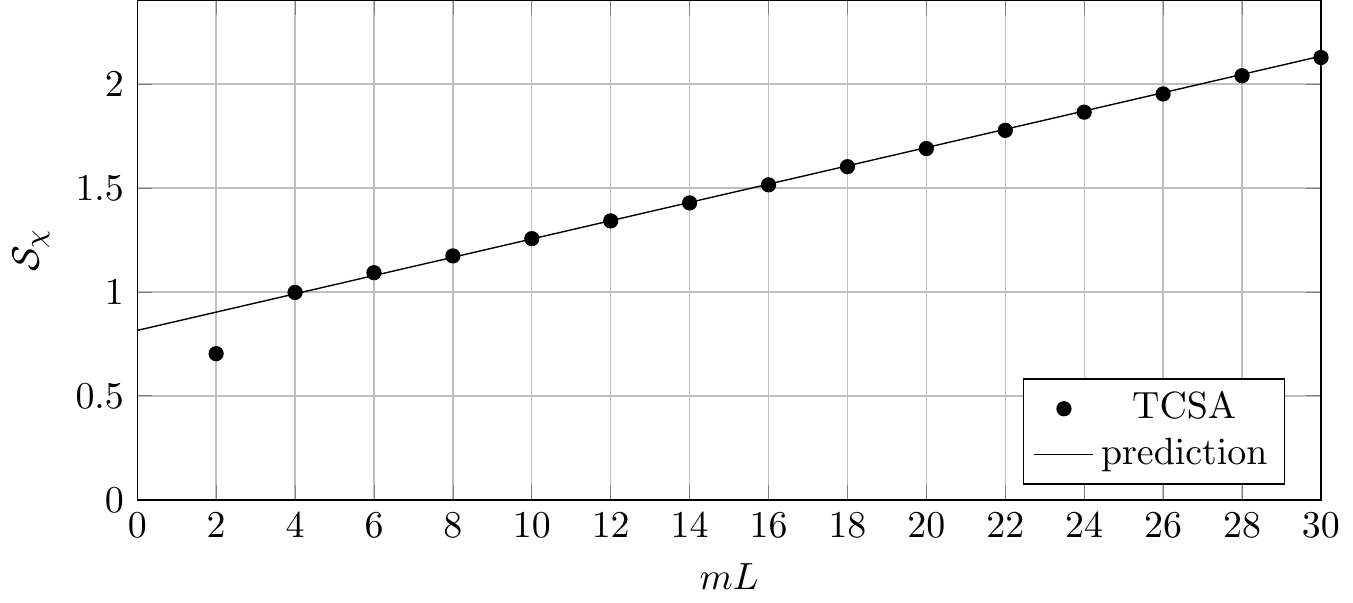}
 \caption{Chiral entanglement entropy in TIFT+$\sigma$. TCSA (dots) against eq.~\eqref{eq:LREE_final} with the corresponding value of $\tau^*$ (see Table~\ref{tab:taus}). TCSA data are extrapolated to infinite cut-off.}
 \label{fig:LREE_m45_3o80}
\end{figure}

\begin{figure}
 \centering
 \includegraphics{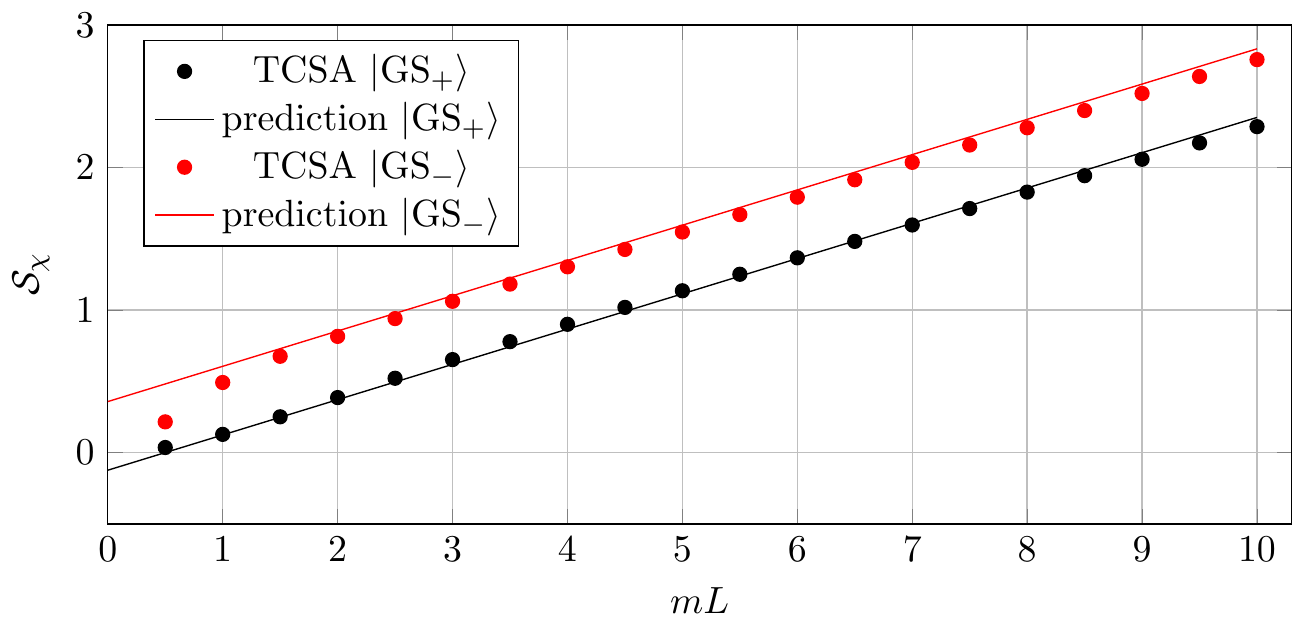}
 \caption{Chiral entanglement entropy in TIFT+$\sigma'$. TCSA data are extrapolated to infinite cutoff.}
 \label{fig:LREE_m45_7o16}
\end{figure}

\begin{itemize}
\item $\sigma$ perturbation: Cardy's Ansatz predicts $|a^*\rangle=|\widetilde{3/2}\rangle$ with the $\tau^*=2.0844592$ (cf. Table~\ref{tab:taus}). The predicted chiral entanglement is compared to the TCSA reuslt in Figure~\ref{fig:LREE_m45_3o80} which shows an excellent agreement. The universal constants $\gamma$ and the slope $\mathcal{B}$ obtained from a linear fit of the TCSA data agrees with the theoretical prediction up to four and three digits respectively as shown in Table~\ref{tab:num_EE_slope_inter}.

\item $\varepsilon$ perturbation: for $t>0$ there is a unique ground state corresponding to $|a^*\rangle=|\widetilde{7/16}\rangle$, while for $t<0$ the two degenerate ground states are $|\text{tNS}\rangle$ and $|\text{tR}\rangle$ given in \eqref{tNS} and \eqref{tR}. The theoretical prediction and the TCSA results for the chiral entanglement entropy are shown in Figure~\ref{fig:LREE_m45_1o10}, and the extrapolated data are in good agreement with the predictions for each case. 

\item $\sigma'$ perturbation: for $h'>0$ there are two ground states $|\text{GS}_+\rangle$ \eqref{dc1} and $|\text{GS}_-\rangle$ \eqref{dc2} which are degenerate in infinite volume. The theoretical prediction and the TCSA results for the chiral entanglement entropy are compared in Figure\ref{fig:LREE_m45_7o16}. For $h'<0$ there are again two ground states, but their expressions differ from the $h'>0$ result (\ref{dc1},\ref{dc2}) by flipping sign of the $|\sigma'\rangle\rangle$/$|\sigma\rangle\rangle$ terms, respectively, as discussed in Section~\ref{sec:exmp}. This sign change has no effect neither on the slope $\mathcal{B}$ nor the intercept $\gamma$, the prediction is the same for both signs, while the overlaps obtained from TCSA are also the same up to a sign flip that has no effect on the chiral entanglement entropy. 

\end{itemize}

\subsubsection{Detailed numerical comparison}
Beyond the graphical comparison presented so far, we also extracted the values of $\mathcal{B}$ and $\gamma$ by fitting the volume dependence of the chiral entanglement entropy obtained from TCSA with a linear function. For any quantity determined from TCSA there is a so-called scaling regime where finite size effects (which decrease with the volume) are of the same order of magnitude as truncation effects (which increase with the volume). For the entanglement entropy, since the predicted behaviour in large volume is linear, the scaling regime can be found by examining the numerically computed second derivative and choosing an interval in $mL$ where it is sufficiently small. This cannot be done completely automatically, as sometimes a blind search would find the minimum of the second derivative in a range that is clearly at too small volume still dominated by finite size effects, and can be aided by the graphical comparison between prediction and TCSA data discussed above. The detailed numerical matching of the predicted  to the results obtained from TCSA is presented in Table~\ref{tab:num_EE_slope_inter}.

\begin{table}[h!]
\centering
\begin{tabular}{|c|c|c|c|c|c|}
\hline 
Perturbed CFT with  & $\gamma$ theory & $\gamma$ TCSA & $\mathcal{B}$ theory & $\mathcal{B}$ TCSA & TCSA fit  \\ 
ground state& & & & & range in $mL$ \\
\hline 
IFT+$\sigma$, $h>0$  & $-0.51986$ & $-0.5168(20)$ & $0.032774$ & $0.032653(49)$ & $20-60$\\ 
\hline 
IFT+$\varepsilon$, $t>0$ & $0$ & $0.0031(29)$ & $0.25$ & $0.24756(44)$ & $5.5-8.5$\\
\hline
IFT+$\varepsilon$, $t<0$, $|\text{NS}\rangle$ & $0$ & $0.0031(29)$ & $0.25$ & $0.24756(44)$ & $5.5-8.5$\\
\hline
IFT+$\varepsilon$, $t<0$, $|\text{R}\rangle$ & $0.346574$ & $0.32196(35)$ & $0.25$ & $0.244009(48)$ & $6-8.5$\\
\hline
TIFT+$\sigma$, $h>0$ & $-0.814618$ & $-0.81476(22)$ & $0.043959$ & $0.043720(10)$ & $18-28$\\
\hline
TIFT+$\varepsilon$, $t>0$ & $-0.294757$ & $-0.3000(12)$ & $0.061731$ & $0.061546(35)$ & $30-40$\\
\hline
TIFT+$\varepsilon$, $t<0$, $|\text{tNS}\rangle$ & $-0.294757$ & $-0.3000(12)$ & $0.061731$ & $0.061545(35)$ & $30-40$\\
\hline
TIFT+$\varepsilon$, $t<0$, $|\text{tR}\rangle$ & $0.051816$ & $0.0577(14)$ & $0.061731$ & $0.06154(36)$ & $32-48$\\
\hline
TIFT+$\sigma'$, $h'>0$, $|\text{GS}_+\rangle$ & $0.123105$ & $0.101(11)$ & $0.247372$ & $0.2491(30)$ & $2-5$\\
\hline
TIFT+$\sigma'$, $h'>0$, $|\text{GS}_-\rangle$ & $-0.358107$ & $-0.3304(15)$ & $0.247372$ & $0.24330(43)$ & $2-5$\\
\hline

\end{tabular} 
\caption{Summary of numerical results on the chiral entanglement entropies in different models. Theoretical predictions for universal constants and slopes are calculated with the corresponding $\tau^*$ and central charge. Results from TCSA are results of linear fit the extrapolated data in the indicated volume region. The parentheses show the uncertainty of the last digits resulting from the fits. Note that this does not account for the total error budget, since it does not include finite size effects, and also residual truncation effects remaining after extrapolation.}
 \label{tab:num_EE_slope_inter}
\end{table}

\section{Conclusions}
\label{sec:conc}
In this paper we analysed entanglement production among UV chiral degrees of freedom along a  massive renormalisation group flow in $1+1$ dimensions for theories with $\mathbb Z_2$ invariant fixed points, belonging to the $A$-series of Minimal Models and perturbed by a relevant scalar field.  We studied the reduced density matrix, obtained from the finite volume ground state projector after tracing out one chiral sector of the Hilbert space, and showed that contrary to real space entanglement, chiral entanglement is finite when the short-distance cut-off is sent to zero.
Moreover, in the large volume limit, the chiral entanglement entropy grows linearly with the system size and contains a subleading constant term $\gamma$, which was argued to be a universal property of the RG flow and can characterise uniquely the QFT ground state also in presence of degeneracies. We considered examples in Ising Field Theory and Tricritical Ising Field Theory. In the latter theory, we also pointed out the existence of a two-fold ground state degeneracy in presence of a $\mathbb Z_2$ odd perturbation of weight $7/8$.  Interestingly, in such a case, the QFT Hilbert space splits into the same two  sectors discussed in~\cite{FTLTKWF} when describing lattice realisations of Fibonacci anyons.   

For an analytic determination of chiral entanglement we employed a recently proposed variational Ansatz~\cite{Cardy17} for the QFT ground state in terms of smeared conformal boundary states. We tested the  variational Ansatz extensively with TCSA, and our detailed numerical analysis of the energy and the overlaps suggests that the variational Ansatz for the QFT ground state is exact in the infinite volume limit. We then computed both the volume density and the constant term of the chiral entanglement from the Ansatz and demonstrated that it matches excellently the TCSA data, which also provides a further validation of Cardy's Ansatz.

There are several interesting directions for future investigation. Among them, we could mention: the extension of the present investigation to non-diagonal theories, such as e.g. the $\mathbb Z_3$ invariant Potts Field Theory~\cite{ZamPotts} or to massive deformations of free bosonic theories, such as e.g. the Sine-Gordon model. The behaviour of the chiral entanglement and its universal features along a massless RG flow~\cite{Zam_massless_flow} is an important challenge as well; in this case a variational Ansatz for the ground state wave function is currently missing.

It would also be interesting  to extract the coefficient $\gamma$ in Eq. \eqref{ent_sum} (cf. also~\eqref{gamma_top}) directly from   lattice  calculations, either analytically or numerically. This requires the construction of the conformal basis~\cite{KS} on the lattice, and also the determination of the ground state overlaps. Both are challenging tasks; nevertheless, some progress has been made recently in~\cite{MV}. We hope our results could trigger more activity in this direction.

There is also a number of potential applications to quantum quenches in field theory since Cardy's Ansatz determines the ground state of a massive quantum field theory, and so it can be used as a starting point to study the time evolution. There has been considerable progress in both numerical \cite{2016NuPhB.911..805R,2017PhLB..771..539H,2018ScPP....5...27H,2018PhRvL.121k0402K,2018arXiv180906789H} and analytical approaches \cite{2012JSMTE..04..017S,2014JSMTE..10..035B,2017JSMTE..10.3106C,2014JPhA...47N2001D,2017JPhA...50h4004D,2016PhRvE..93f2101K,2016NuPhB.902..508H,2018JHEP...08..170H}. However, finding the overlaps of the initial state with the post-quench energy levels and characterising the time evolution especially for long times is still largely an open problem. Chiral entanglement also looks interesting as a potential tool to analyse and understand the temporal dynamics.
 
\section*{Acknowledgments}
We are grateful to F. Essler for a useful discussion, and also to J. Dubail and H. Katsura for discussions and for drawing our attention to the papers~\cite{DubailReadRezayi} and~\cite{KL}. M.L. was supported by the Brazilian Ministry of Education. G.T. was supported by the National Research Development and Innovation Office (NKFIH) under a K-2016 grant no. 119204 by the Quantum Technology National Excellence Program (Project No. 2017-1.2.1-NKP-2017- 00001), and also by the BME-Nanotechnology FIKP grant of EMMI (BME FIKP-NAT).

\appendix
\section{CFT and TCSA on a cylinder}
\label{appA}
We briefly recall some relevant notions regarding diagonal CFTs on a cylinder to set up our conventions. The Euclidean space-time cylinder of circumference $L$ is parameterised by the Euclidean time $x$ and the spatial coordinate $y$ with the identification $y\equiv y+L$, which can be mapped onto the conformal plane by the exponential function
\begin{equation}
z=\exp\left\{\frac{2\pi}{L}\left(x+iy\right)\right\}\quad,\quad
\bar{z}=\exp\left\{\frac{2\pi}{L}\left(x-iy\right)\right\}\,.
\end{equation}
The conformal Hamiltonian on the cylinder can be writen in terms of the left/right moving 
($l$/$r$) Virasoro generators on the plane as
\begin{equation}
\label{hamCFT}
H_{\text{CFT}}=\frac{2\pi}{L}\left(L_0^l+L_0^r-\frac{c}{12}\right)\,,
\end{equation}
and the eigenvalues of $L_0^l+L_0^r$ are called the scaling dimensions. For a diagonal CFT the states in the conformal basis are of the form 
\begin{equation}
|h,N,k\rangle_l\otimes|h,N',k'\rangle_r\,,
\end{equation}
where $h$ is the highest weight of the left and right component, $N$ and $N'$ are the left/right descendant levels and the integers $k$ and $k'$ label an orthonormal basis at the given level of left/right moving degrees of freedom which satisfy 
\begin{equation}
L_0^{l,r}|h,N,k\rangle_{l,r}=(h+N)|h,N,k\rangle_{l,r}\,.
\end{equation}
The conserved (spatial) momentum operator on the cylinder $P$ is given by 
\begin{equation}
\label{PCFT}
P=\frac{2\pi}{L}\left(L_0^l-L_0^r\right).
\end{equation}
where the eigenvalues of $L_0^l-L_0^r$ give the conformal spin. All states in the Hilbert space have integer conformal spin given by the difference $N-N'$ of the left and right descendant levels. For the minimal models considered in the main text, the set of allowed values for $h$ and consequently for the  conformal dimension at the UV fixed point is finite. The perturbing operator is one of the spin-$0$ fields in the CFT spectrum, which is supposed to be relevant and therefore it is a primary field.

Using the exponential mapping, the matrix elements of the Hamiltonian can be written as
\begin{equation}
\label{matH}
    H_{ij}=\delta_{s_i,s_j}\frac{2\pi}{L}\left(2h_i+N_i^l+N_i^r-\frac{c}{12}+\lambda\frac{L^{2-2h}}{(2\pi)^{1-2h}}\mathrm{B}_{ij}\right)\,,
\end{equation}
where $s_i, s_j$ are the conformal spin of the states, $h$ is the conformal weight of the perturbing field $\phi$ (satisfying $h<1$, with the scaling dimension of $\phi$ given by $\Delta=2h$), and
\begin{equation}
\label{matB}
    \mathrm{B}_{ij}=\langle i |\phi(1,1)|j \rangle_{pl}
\end{equation}
is the matrix element on the plane which can be evaluated in terms of the known structure constants of the CFT~\cite{DF} using the conformal Ward identities. 

The coupling constant is a dimensionful quantity and can be represented in terms of a mass scale $m$ as 
\begin{equation}
 \lambda=\kappa m^{2-2h}
\label{massgap}\end{equation}
where $\kappa$ is a dimensionless number. The standard choice for $m$ is provided by the massgap i.e. the mass of the lowest lying particle, and for integrable models the values of $\kappa$ resulting in this particular choice of units are exactly known; for the models used in this paper they can be found in \cite{Fat94} and are listed in Table \ref{tab:taus}. For the non-integrable perturbation of the tricritical Ising model with the leading magnetization operator $\sigma$, we used $\kappa=0.1$ to define a mass scale $m$ which for this particular case is just an arbitrary choice of units.

Substituting \eqref{massgap} into \eqref{matH} shows that the finite volume spectrum can be considered as a function of the dimensionless variable $mL$ when the energies are measured in units of $m$, and therefore a CFT perturbed by a single relevant field has no free parameters apart from the choice of the perturbing field, and the overall sign of the coupling when the perturbing field is even.

Due to translational invariance the Hilbert space can be split into different momentum sectors. We restrict our attention to the zero-momentum sector since we are interested in the ground state. To set up the TCSA we truncate the Hilbert space by introducing a cut-off $\Lambda$ in the descendant level $N$ and keep only states with $N\leq\Lambda$, resulting in a restriction of the Hamiltonian to a finite dimensional matrix. The dimensions of the truncated zero momentum sectors in the Ising and tricritical Ising models are given in Table~\ref{tab:tcsadims}.

States from the TCSA at cut-off $\Lambda$ and in volume $L$ can be written in the following way
\begin{equation}
|\Psi_{\Lambda}(L)\rangle = \sum_{a,b} v_{ab,\Lambda}(L) |a\rangle_l\otimes|b\rangle_r
\end{equation}
Since the ground state is in the zero momentum sector $v_{ab,\Lambda}(L)=\delta_{h_a,h_b}\delta_{N_a,N_b} \omega_{ab}$, therefore the density matrix is block-diagonal formed by blocks from the same module and descendant level. The partial trace necessary to obtain the reduced chiral density matrix can be carried out separately in each block, and the chiral entanglement spectrum can be obtained by diagonalizing the reduced blocks separately. From the entanglement spectrum one can calculate the chiral R\'enyi entropies $S^n_{\chi,\Lambda}(L)$ and the chiral entangement entropy $S_{\chi,\Lambda}(L)$ at a given truncation level $\Lambda$ as a function of the volume $L$.

\begin{table}
    \centering
    \begin{tabular}{|c|ccc||c|cccccc|}
    \hline
        \multicolumn{4}{|c||}{Ising} & \multicolumn{7}{c|}{Tricritical Ising} \\
        \hline
        $\Lambda$ & $1$ & $\sigma$ & $\varepsilon$ & $\Lambda$ & $1$ & $\sigma$ & $\varepsilon$ & $\sigma'$ & $\varepsilon'$ & $\varepsilon''$\\
        \hline
        $15$ & $1037$ & $2167$ & $1272$ & $9$ & $156$ & $877$ & $449$ & $428$ & $606$ & $389$\\
        $16$ & $1566$ & $3191$ & $1897$ & $10$ & $300$ & $1553$ & $810$ & $752$ & $1090$ & $678$\\
        $17$ & $2242$ & $4635$ & $2738$ & $11$ & $496$ & $2709$ & $1386$ & $1281$ & $1819$ & $1119$\\
        $18$ & $3331$ & $6751$ & $3963$ & $12$ & $896$ & $4645$ & $2410$ & $2181$ & $3115$ & $1903$\\
        $19$ & $4700$ & $9667$ & $5644$ & $13$ & $1425$ & $7781$ & $4010$ & $3625$ & $5140$ & $3059$\\
        $20$ & $6816$ & $13763$ & $8045$ & $14$ & $2449$ & $12822$ & $6611$ & $5929$ & $8504$ & $4995$\\
        \hline
    \end{tabular}
    \caption{Dimensions of the truncated modules at different cut-offs in the critical and in the tricritical Ising field theory}
    \label{tab:tcsadims}
\end{table}

\section{Cut-off extrapolation and finite volume deviations from the Ansatz}
\label{appB}

\begin{figure}
 \centering
 \includegraphics{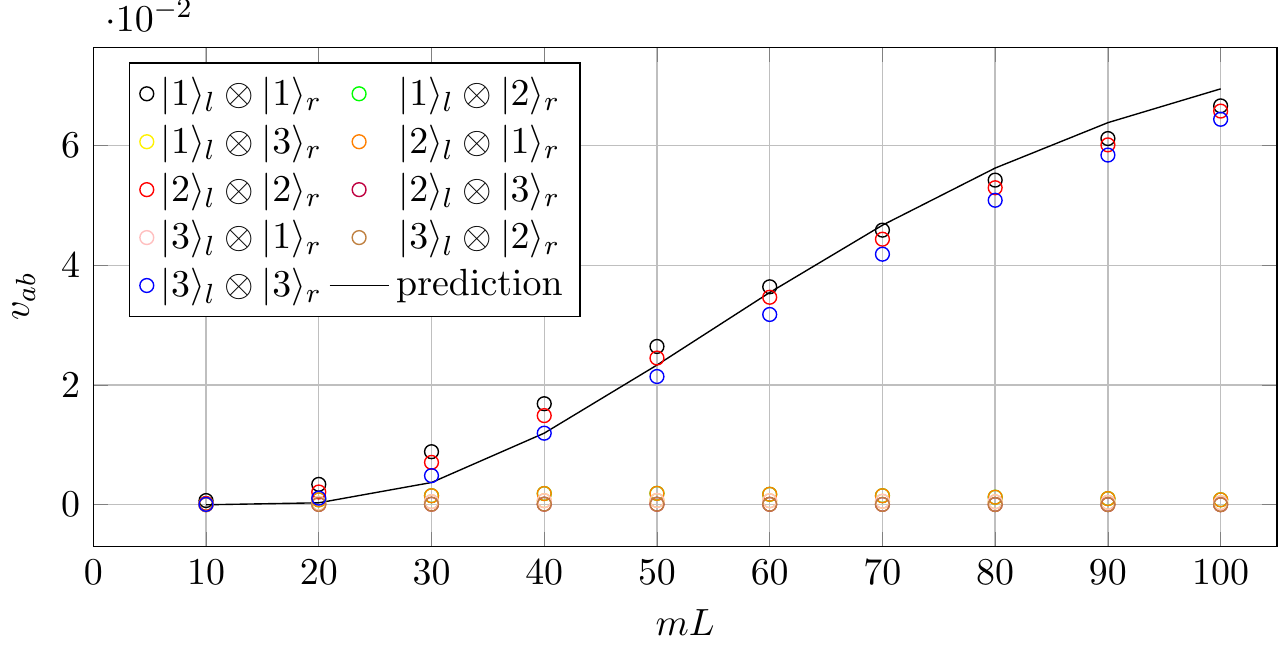}
 \caption{Behaviour of the overlaps in $E_8$ field theory with increasing volume. The data indicated by circles are the extrapolated overlaps from TCSA at descendent level $6$ in the identity module of the critical Ising field theory. There are $3$ states in each chiral sector (labelled by $|1\rangle_{l,r}$, $|2\rangle_{l,r}$ and $|3\rangle_{l,r}$), therefore there are $9$ states in the full Hilbert space in the zero momentum sector at this level. The Ansatz predicts a vanishing overlap when the left and right chiral components differ, while the continuous line shows the prediction for the diagonal overlaps calculated using~\eqref{overgs}}
 \label{fig:maverick}
\end{figure}

In this Appendix we describe the cut-off extrapolation procedure using the TCSA renormalization group approach developed in \cite{FGPTW,KA,GW,LT,Rychkov1,Rychkov2}. 
The full Hilbert space can be split as $\mathcal{H}=\mathcal{H}_l \oplus \mathcal{H}_h$ where $\mathcal{H}_l$ and $\mathcal{H}_h$ the low and high energy subspaces respectively given in terms. The full Hamiltonian can be written in a block diagonal form as
\begin{equation}
    H=\left(
    \begin{array}{cc}
     H_{ll} & H_{hl} \cr
     H_{lh} & H_{hh}
    \end{array}
    \right)\,.
\end{equation}
In the conformal basis the off-diagonal parts only contain the perturbing field and there dependence of the coupling can be written as 
\begin{equation}
 H_{hl}=\lambda V_{hl}\quad,\quad H_{lh}=\lambda V_{lh}\,.
\end{equation}
Diagonalizing the truncated Hamiltonian $H_{ll}$ results in energy levels and eigenstates  $\{E_{\Lambda},|\Psi_{\Lambda}\rangle\}$. Since the perturbation is a relevant operator, the effect of the high-energy subspace can be taken into account perturbatively, which results in 
\begin{align}
    E_{\infty} & =  E_{\Lambda}-\lambda^2 \sum_{n\in\mathcal{H}_h}\frac{\langle \Psi_{\Lambda}|V_{lh}|n\rangle \langle n|V_{hl} | \Psi_{\Lambda}\rangle}{E_n-E_{\Lambda}}+O(\lambda^3) \nonumber\\
    |\Psi_{\infty}\rangle & =  E_{\Lambda}+\lambda^2 \sum_{n,m\in\mathcal{H}_h}|m\rangle \frac{\langle m|V_{hh}|n\rangle \langle n|V_{hl} | \Psi_{\Lambda}\rangle}{(E_n-E_{\Lambda})(E_m-E_{\Lambda})} 
    \nonumber\\
    & -\frac{\lambda^2}{2}|\Psi_{\Lambda}\rangle\sum_{n\in\mathcal{H}_h} \frac{\langle \Psi_{\Lambda}|V_{lh}|n\rangle \langle n|V_{hl} | \Psi_{\Lambda}\rangle}{(E_n-E_{\Lambda})^2}+O(\lambda^3)
\label{cutofftheory}
\end{align}
to the first nontrivial order, where $E_n$ scales linearly with $\Lambda$. The dependence of the perturbing matrix elements on the cutoff can be evaluated with the result~\cite{GW, LT}
\begin{equation}
    \langle \Psi_{\Lambda}|V_{lh}|n\rangle \langle n|V_{hl} | \Psi_{\Lambda}\rangle= \sum_{a\in\phi\times\phi}\Lambda^{2h_a-1}\left(A_{a,0}+\frac{A_{a,1}}{\Lambda}+\frac{A_{a,2}}{\Lambda^2}+\dots\right)
\end{equation}
where $h_a=2h_{\phi}-h_a$ and the summation is over the fields which can be found in the OPE of the perturbation with itself. Therefore to leading order the cut-off dependence of energy levels can be extrapolated with a function of the form
\begin{equation}
    E_{\Lambda}=E_{\infty}+A\Lambda^{e_1}+B\Lambda^{e_2}
\end{equation}
where the leading exponent is always $e_1=4h-2$ corresponding to the identity operator in the OPE, while the next-to-leading $e_2$ depends on the particular model and perturbing operator under consideration.
\begin{table}
\centering
\begin{tabular}{|c|c|c|c|c|c|c|}
\hline 
 Model & OPE & 1 & 2 & 3 & 4 & To fit \\ 
\hline 
IFT & $\sigma\times\sigma=1+\varepsilon$ & $-\frac{11}{4}$
 & $-\frac{15}{4}$ 
&  &  & $-\frac{11}{4},-\frac{15}{4}$
 \\
\hline 
IFT & $\varepsilon\times\varepsilon=1$ & $-1$ &  &  &  & $-1,-2$ \\
\hline 
TIFT & $\sigma\times\sigma=1+\varepsilon+\varepsilon'+\varepsilon''$ & $-\frac{57}{20}$
 & $-\frac{61}{20}$
 & $-\frac{81}{20}$
 & $-\frac{117}{20}$
 & $-\frac{57}{20},-\frac{61}{20}$ \\
\hline 
TIFT & $\varepsilon\times\varepsilon=1+\varepsilon'$ & $-\frac{13}{5}$
 & $-\frac{19}{5}$
 &  &  & $-\frac{13}{5},-\frac{18}{5}$ \\
\hline 
TIFT & $\sigma'\times\sigma'=1+\varepsilon''$ & $-\frac{5}{4}$
 & $-\frac{17}{4}$
 &  &  & $-\frac{5}{4},-\frac{9}{4}$\\
\hline 
\end{tabular}
\caption{Fitting exponents for overlaps and chiral entanglement entropy. The columns $1$-$4$ give the leading exponent coming from the corresponding term of the OPE. Note that each one is accompanied by sub-leading exponents resulting from $1/\Lambda$ corrections, and in some cases such as TIFT perturbed by $\epsilon$ or by $\sigma'$, a sub-leading exponent from the identity channel is more relevant than a leading one from another channel. For the thermal perturbation of the Ising model the OPE contains only the identity channel, so all exponents come from the identity contribution and are given by negative integers. The last column lists the exponents we used for cut-off extrapolation.}
\label{tab:exponents}
\end{table}
From \eqref{cutofftheory} the overlaps with the low energy CFT states can be extrapolated with the same functional form with exponents shifted by $+1$, therefore the leading one is given by $e_1=4h-3$. The exponents necessary for our calculations are summarized in  Table~\ref{tab:exponents}, while Figure~\ref{fig:maverick} shows the results of extrapolation in the $E_8$ scattering theory for overlaps with conformal states at level $6$ in the identity module using cut-offs $10\leq\Lambda\leq 20$. Besides demonstrating the extrapolation procedure it also demonstrates that the deviations of the overlaps from prediction of eq.~\eqref{overgs} and the overlaps converge to the predicted value when the volume is increased. 

\begin{figure}
 \centering
 \includegraphics{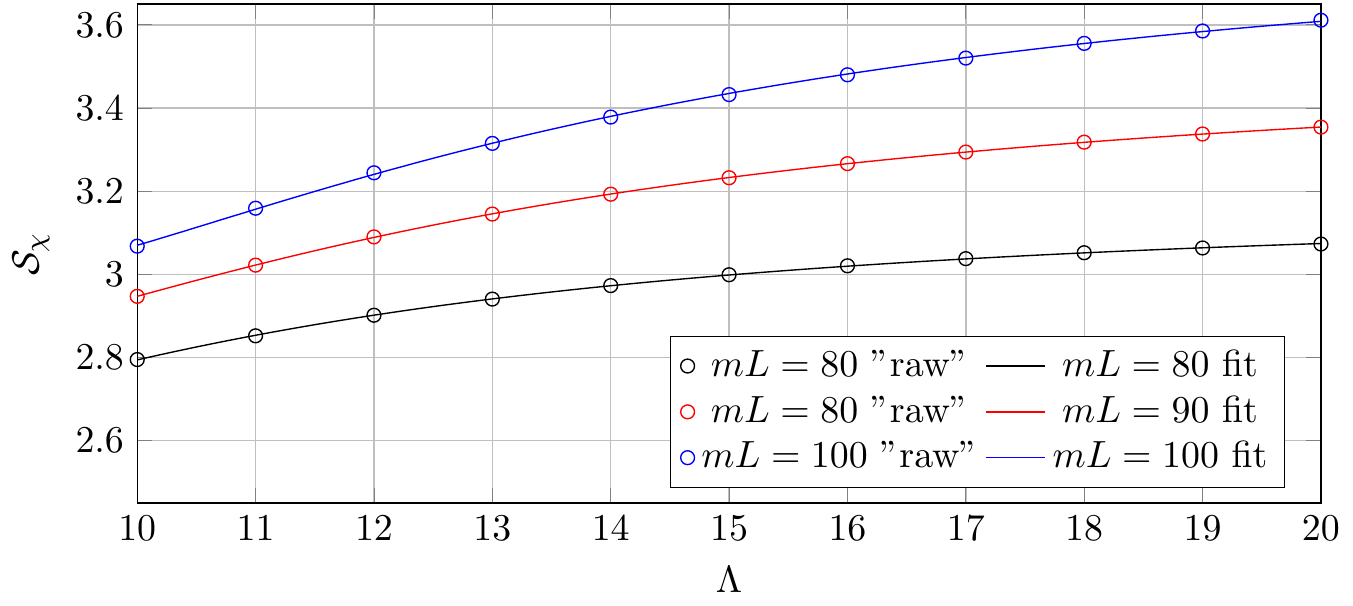}
 \caption{Cut-off extrapolation of the chiral entanglement in the $E_8$ field theory for different volumes.}
 \label{fig:lreextra}
\end{figure}

For the chiral entanglement entropy we assumed that the TCSA cut-off is high enough so that the overlaps only differ from their exact value by a small amount, therefore the cut-off dependence of the chiral entanglement entropy can be considered as a linear function of cut-off dependence of the overlaps. Therefore the chiral entanglement entropy was extrapolated with the same exponents as the overlaps, and Figure~\ref{fig:lreextra} demonstrates that the proposed extrapolating functions indeed provide an excellent fit.

\bibliographystyle{JHEP}

\providecommand{\href}[2]{#2}\begingroup\raggedright\endgroup

\end{document}